\documentclass[fleqn,usenatbib]{mnras}

\usepackage{newtxtext,newtxmath}

\usepackage{xcolor}
{}
\usepackage{cancel}
\usepackage{booktabs}
\usepackage{threeparttable}
\usepackage{soul,xcolor}
\setstcolor{red}
\usepackage{lineno}
\usepackage{bm}

\usepackage[T1]{fontenc}

\DeclareRobustCommand{\VAN}[3]{#2}
\let\VANthebibliography\thebibliography
\def\thebibliography{\DeclareRobustCommand{\VAN}[3]{##3}\VANthebibliography}

\usepackage{graphicx}	
\usepackage{amsmath}	
\usepackage{siunitx}
\usepackage{threeparttable}
\usepackage{comment}
\usepackage{natbib}

\newcommand{\ceo}{\rm C^{\rm 18}O}
\newcommand{\tco}{\rm ^{\rm 13}CO}

\title[L1616 gas kinematics]{Investigating the Kinematics of Molecular Gas in Cometary Globule L1616}

\author[Porel et al.]{
Puja Porel,$^{1}$$^{,}$$^{2}$ \thanks{E-mail: pujaporel11@gmail.com}
Archana Soam,$^{1}$
Janik Karoly,$^{3}$
Eun Jung Chung,$^{4}$
Chang Won Lee$^{4}$
\\
$^{1}$Indian Institute of Astrophysics, II Block, Koramangala, Bengaluru 560034, India \\
$^{2}$ Pondicherry University, R.V. Nagar, Kalapet, Puducherry, 605014, India\\
$^{3}$Jeremiah Horrocks Institute, University of Central Lancashire, Preston PR1 2HE, UK\\
$^{4}$Korea Astronomy and Space Science Institute (KASI), 776 Daedeokdae-ro, Yuseong-gu, Daejeon 34055, Republic of Korea\\}

\date{Accepted XXX. Received YYY; in original form ZZZ}

\begin{document}
\label{firstpage}
\pagerange{\pageref{firstpage}--\pageref{lastpage}}
\maketitle

\begin{abstract}
LDN 1616 is a cometary globule located approximately $8^\circ$ west of the Orion OB1 associations. The massive OB stars in the Orion belt region act as catalysts, triggering the star formation activity observed in the L1616 region which is a photodissociation region (PDR). This paper provides an in-depth analysis of gas kinematics within the L1616 PDR, leveraging the Heterodyne Array Receiver Program (HARP) on the James Clerk Maxwell Telescope (JCMT) to observe $\mathrm{^{13}CO}$ and $\mathrm{C^{18}O}$ $J = 3 \rightarrow 2$ emissions. Employing the Clumpfind algorithm on the C$^{18}$O emission data, we identify three distinct clumps within this PDR. For each of these clumps, we derive key physical parameters, including the mean kinetic temperature, optical depth, and velocity dispersion.
In addition, we compute the non-thermal velocity dispersion and Mach number, providing critical insights into the turbulent dynamics of the gas. Comprehensive evaluation of mass, including virial and energy budget evaluations, are conducted to assess the gravitational stability and star-forming potential of the identified clumps. While previous studies have proposed that radiation-driven implosion (RDI) is the dominant mechanism initiating star formation in LDN 1616, our results suggest that the clumps may represent pre-existing substructures within the PDR. This interpretation is supported by our estimation of a relatively low interstellar radiation field ($G_0$), which, although insufficient to form clumps independently, may enhance gravitational instability through additional compression. Thus, our findings offer a more nuanced perspective on the role of RDI, highlighting its capacity to trigger star formation by amplifying the instability of pre-existing clumpy structures in PDRs like LDN 1616.

\end{abstract}




\section{Introduction}
Star formation within molecular clouds is significantly influenced by the dynamic interactions of winds and H\,\textsc{ii} regions \citep{SHORE2003727}. These processes can either dissipate the cloud through radiative and mechanical heating or trigger star formation through phenomena such as champagne flows. The latter occurs when the breakout of a shock wave from a hot H\,\textsc{ii} region surrounding OB stars causes a rapid outflow of material. This outflow drives a shock into the molecular cloud, compressing its dense core and potentially initiating local gravitational collapse, thereby propagating star formation.

Cometary globules (CGs), first identified in 1976 \citep{hawarden1976cometary}, are a class of interstellar clouds distinguished by their comet-like morphology, featuring compact, dusty heads and elongated faintly luminous tails. Unlike typical dark clouds, CGs are isolated neutral globules surrounded by hot ionized media. Two primary models have been proposed to explain their formation: \cite{1983MNRAS.203..215B} suggested that CGs were initially spherical clouds shocked by supernova blast waves, while \cite{Reipurth_2021} posited that ultraviolet (UV) radiation impinging on neutral clouds within a clumpy interstellar medium shapes these globules. Discriminating between these models necessitates detailed knowledge of the mass, density, temperature, and velocity distribution of the globules, which can be derived from molecular spectral line measurements at millimeter wavelengths.

\begin{figure*}
\begin{center}
\resizebox{18.0cm}{10cm}{\includegraphics{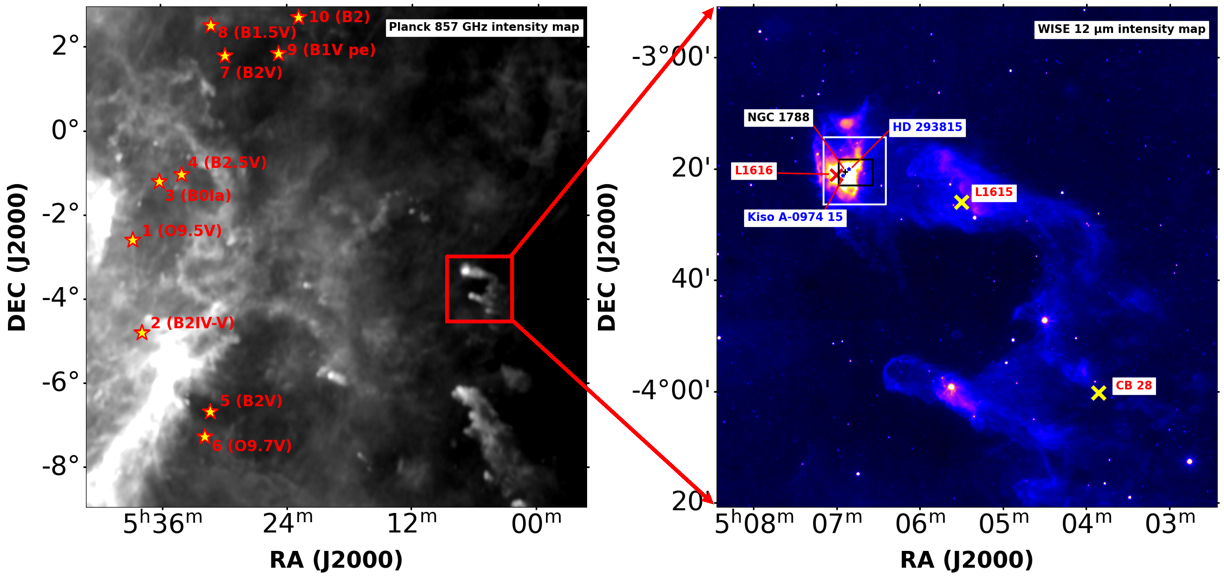}}
\caption{\textbf{Left:} The Planck 857 GHz intensity map covers an area of $12^\circ \times 12^\circ$ with an angular resolution of $5^\prime$, showcasing the positions of the prominent Orion stars, annotated with their spectral types. The numerical identifiers assigned to these stars correspond to those listed in Table 1 of \citet{saha2022magnetic}, who adopted the stellar sample from the earlier work of \citet{ramesh1995study}. A red box, approximately $1.5^\circ \times 1.5^\circ$ in size, highlights the locations of L1616, L1615, and CB 28.
\textbf{Right:} A zoomed-in view of the red box from the left panel reveals the WISE 12 $\mu$m intensity map, spanning $1.5^\circ \times 1.5^\circ$ with an angular resolution of 6.5".  Red and Yellow crosses mark the positions of the centers of the clouds L1616, L1615, and the CB 28 Bok globule. The white and black boxes delineate the regions surveyed in $^{13}$CO (3-2) and C$^{18}$O (3-2) emissions within the L1616 photodissociation region. The black plus sign and blue stars denote the locations of the reflection nebula NGC 1788 and the intermediate-mass stars Kiso A-0974 15 and HD 293815, respectively.
}\label{Fig: Planck 857 GHz image and WISE 12 micron image}
\end{center}
\end{figure*}

The Lynds Dark Nebula LDN 1616 \citep{ramesh1995study}, located approximately $8^\circ$ west of the Orion OB1 associations \citep{saha2022magnetic, maddalena1986large, stanke2002triggered, alcala2004multi, lee2005triggered, lee2007triggered, caballero2008low}, exemplifies such a cometary cloud (see Figure~\ref{Fig: Planck 857 GHz image and WISE 12 micron image}). The distance to the globule is adopted as $384 \pm 5$ pc \citep{saha2022magnetic}. In the left panel of Figure~\ref{Fig: Planck 857 GHz image and WISE 12 micron image}, the positions of L1616, L1615, and CB 28 are delineated within a red box superimposed on the Planck 857 GHz intensity map. The ten ionizing OB stars situated in the Orion molecular cloud are presented with star symbols, which are at the distances of 48 - 86 pc away from L1616, except the star of spectral type B2 being at a distance of 202 pc from L1616 \citep{saha2022magnetic}. In the right panel, the WISE 12 $\mu$m intensity map displays the positions of the L1616 and L1615 clouds, along with the Bok globule CB 28, marked by the red and yellow crosses respectively. The L1616 photodissociation region (PDR) extends about 40$^{\prime}$ in the sky \citep{alcala2004multi}, and it points towards the east and exhibits active star formation \citep{nakano1995survey, stanke2002triggered, alcala2004multi}. The L1616 PDR features the bright reflection nebula NGC 1788, excited by a star cluster \citep{stanke2002triggered}, with notable members including the intermediate-mass stars Kiso A-0974 15 (B3e) and HD 293815 (B9V) \citep{gandolfi2008star, ramesh1995study}. Remarkably, the star formation efficiency within NGC 1788 is around $14\%$, significantly higher than typical reflection nebula-associated clouds \citep{ramesh1995study}. The star formation efficiency (SFE) in the cloud ranges from $7\%$ to $8\%$, aligning with the typical values observed in molecular clouds situated near OB associations \citep{gandolfi2008star}.

For several reasons, studying gas kinematics in molecular line emissions within molecular clouds is crucial for understanding star formation processes. Tracing gas motion through molecular line emissions, particularly from molecules like carbon monoxide (CO), provides velocity information via Doppler shifts, revealing gas dynamics within the cloud. Kinematic observations can pinpoint regions undergoing gravitational collapse, which is essential for identifying protostar formation sites. Additionally, analyzing gas kinematics reveals turbulence levels and scales, which influence star formation rates and the initial mass function of stars. Combined with line intensity measurements, kinematic data enable estimates of gas mass and density, which is crucial for understanding star-forming conditions. Furthermore, gas kinematics elucidate interactions between magnetic fields and molecular gas, which is vital for comprehending cloud support mechanisms and guiding gas flow \citep{Ching_2018}. Kinematic studies also illuminate feedback processes from young stars and supernovae, impacting cloud turbulence and subsequent star formation \citep{10.1093/mnras/stz2368}. Finally, velocity mapping within molecular clouds identifies active or impending star formation regions, characterized by distinct kinematic signatures.

In this study, we investigate the gas kinematics and physical conditions of the L1616 cometary globule to understand the ongoing star formation activity within this PDR. Utilizing molecular emission data from the $\tco$ and $\ceo$ J = 3-2 transitions, observed by the Heterodyne Array Receiver Program (HARP) on the James Clerk Maxwell Telescope (JCMT) at an angular resolution of 15” \citep{buckle2010jcmt}, we provide a detailed analysis. Section~\ref{section: Observations and data reduction} outlines our observations and data reduction procedures. Section~\ref{section: Results} presents the results, followed by an analysis of the physical properties in Section~\ref{section: Analysis}. Section~\ref{section: Discussion} discusses the scientific implications of our findings, and Section~\ref{section: Summary} concludes with a summary of our research on the L1616 cometary globule.

\section{Observations and data reduction}
\label{section: Observations and data reduction}

Observations of $\mathrm{^{13}CO}$ and $\mathrm{C^{18}O}$ in L1616 were conducted September -- November 2023 with the HARP \citep{buckle2010jcmt} on the JCMT under the project code E23BU009 (P.I. Janik Karoly). The observations were taken in weather bands 3 and 4 (0.08 $< \tau_{225 \, \mathrm{GHz}} <$ 0.20). Each observation covers a 10$\arcmin \times$ 10$\arcmin$ field of view (fov) centered on the bright 12\,$\mu$m emission seen in Figure~\ref{Fig: Planck 857 GHz image and WISE 12 micron image}. The original footprint was centered at RA=05:07:00, Dec=-03:21:06 but after preliminary reductions, the western region of L1616 was on the edge of the fov, so we moved the central coordinates to RA=05:06:53, Dec=-03:21:06. The total on-target integration time was $\approx$5.5 hours. The observed velocity resolution for $^{13}$CO (3-2) and C$^{18}$O (3-2) transitions is 0.05\, km\,s$^{-1}$ and the respective rest frequencies are 330.59 GHz and 329.33 GHz.

The observations were conducted with one of the HARP observing modes, the $\mathrm{^{13}CO}$ and $\mathrm{C^{18}O}$ position-switch basket weave raster mapping with 1/4 array scan spacing. HARP has a main beam efficiency of 0.61 at 345\,GHz \citep{buckle2010jcmt}.

The data were reduced using the {\tt oracdr\_acsis} pipeline \citep{2015A&C.....9...40J} and the REDUCE\_SCIENCE\_NARROWLINE recipe. The data were reduced using a 12$\arcsec$ pixel scale to increase the signal-to-noise and sample the JCMT/HARP beam size. The $^{13}$CO and C$^{18}$O data were reduced separately. Upon initial reductions of the data, the HARP receptors H02 and H12 were consistently flagged as ‘bad’ and so were removed from the data reduction process.

\section{Results}
\label{section: Results}

The formation of CO, a simple yet abundant molecule in the interstellar medium, primarily occurs through gas-phase reactions. One notable advantage of using CO isotopologues, such as $\tco$ and $\ceo$, as tracers of molecular clouds is their resilience against destructive reactions due to their strong binding energies \citep{stahler2008formation}. 

The choice of using $\tco$ and $\ceo$ stems from their distinct characteristics. $\tco$ emission, being optically thick compared to $\ceo$, is particularly useful for studying diffuse regions within the cloud. In contrast, C$^{18}$O emission, being optically thin, is better suited for probing denser regions. By leveraging these isotopologues, we aim to thoroughly characterize the gas dynamics and associated physical conditions within the L1616 PDR, thereby gaining deeper insights into the star formation processes occurring within this photodissociation region.

\begin{figure*}
\begin{center}
\resizebox{12.1cm}{11.1cm}{\includegraphics{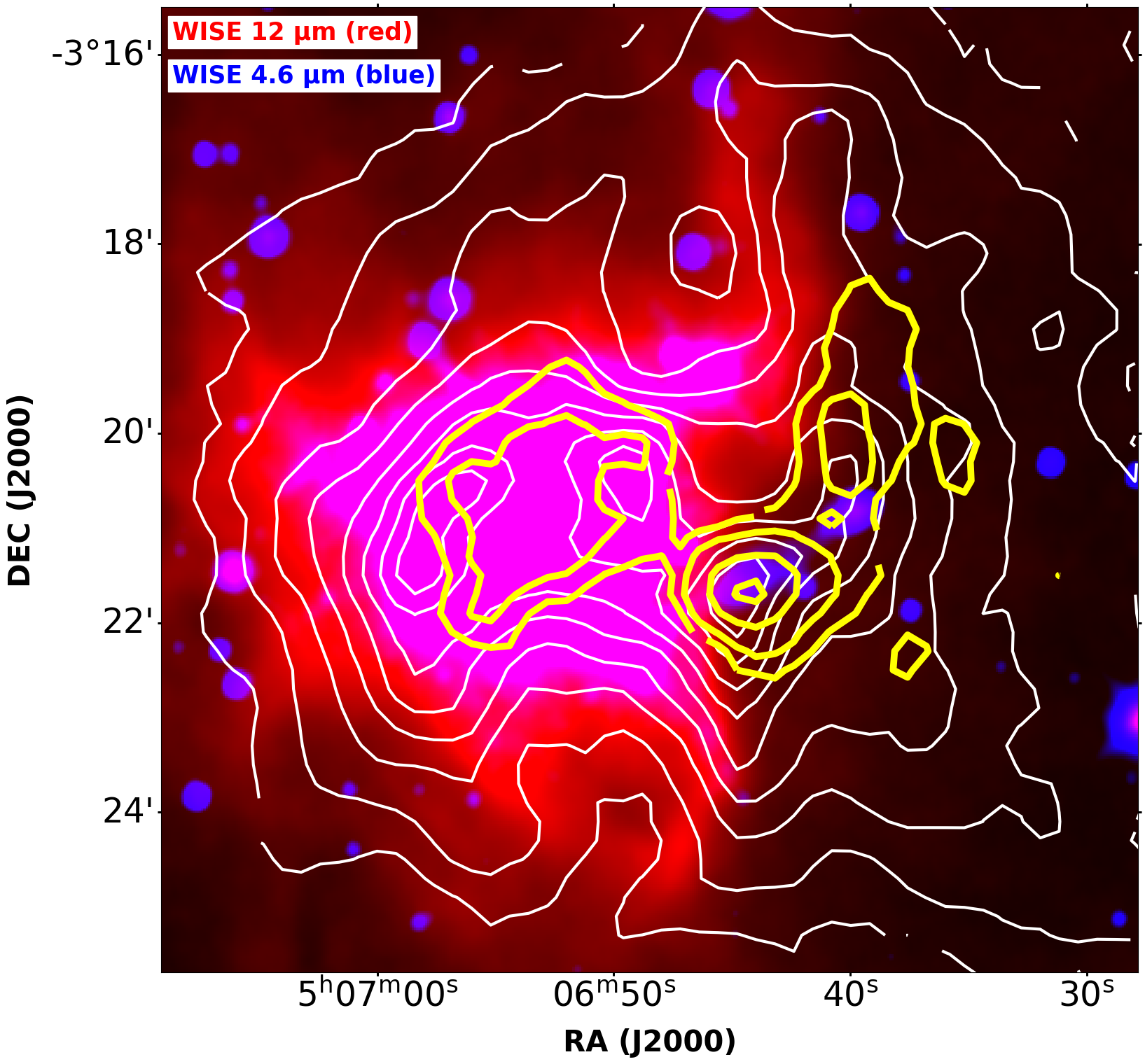}}
\caption{The cometary globule L1616 is vividly portrayed in a WISE two-color image, accentuating its remarkable features at 12 $\mu$m (rendered in red) and 4.6 $\mu$m (in blue), with angular resolutions of approximately 6.5" and 6.4", respectively. The image encompasses an area of approximately $10^\prime \times 10^\prime$. Superimposed on this image are white contours that trace the $^{13}\mathrm{CO}$ (3-2) emission, integrated over a velocity range of 6.35 to 9.12 km s$^{-1}$, with contour levels set at 0.41, 1.38, 2.76, 4.13, 5.51, 6.89, 8.27, 9.64, 11.02, and 12.40 K km s$^{-1}$. These contours are drawn at 3$\sigma$ above the background level, where $\sigma \approx 0.14$ K km s$^{-1}$ represents the standard deviation of the background emission. In addition, yellow contours delineate the C$^{18}$O (3-2) emission, integrated over the same velocity range, with contour levels at 0.55, 0.92, 1.85, 2.77, and 3.69 K km s$^{-1}$. These contours are similarly drawn at 3$\sigma$ above the background, where $\sigma \approx 0.18$ K km s$^{-1}$ is the standard deviation of the background emission. The $^{13}\mathrm{CO}$ and C$^{18}$O emission data are derived from JCMT-HARP observations, boasting an angular resolution of 15".
}\label{Fig: WISE two color (12 micron and 4.6 micron) image map}
\end{center}
\end{figure*}

\subsection{Moment Maps and Spectral Profiles of Molecular Emissions}
\label{section: Moment Maps and Spectral Profiles of Molecular Emissions}

The WISE 12 $\mu$m emission, which traces polycyclic aromatic hydrocarbon (PAH) activity, is contrasted with the WISE 4.6 $\mu$m emission, representative of diffuse dust structures. These emissions are combined into a two-color composite image to accentuate the characteristics of the L1616 PDR, specifically highlighting its dust and PAH contributions, as illustrated in Figure~\ref{Fig: WISE two color (12 micron and 4.6 micron) image map}. The spatial distribution of $\mathrm{^{13}CO}$ (3-2) and $\mathrm{C^{18}O}$ (3-2) emissions within this region are delineated by white and yellow contours, respectively. From this visualization, it becomes evident that the $\mathrm{^{13}CO}$ (3-2) emission envelopes both the WISE 12 $\mu$m and 4.6 $\mu$m emissions, with a more pronounced extension toward the western side compared to other directions. In contrast, the $\mathrm{C^{18}O}$ emission displays a more modest extension to the west, only slightly exceeding the spatial extent of both WISE emissions. While the $\mathrm{^{13}CO}$ and $\mathrm{C^{18}O}$ emissions are visible in the center of the 12 $\mu$m emission, they are more intense next to the western sharp edge in the WISE 12 $\mu$m image (R3 and C2). The more extensive spread of $\mathrm{^{13}CO}$ emission compared to $\mathrm{C^{18}O}$ emission indicates that $^{13}$CO effectively traces the outer, more diffuse regions of the cloud. In contrast, C$^{18}$O serves as a better tracer for the denser, central regions of the cloud.

\begin{figure*}
\begin{center}
\resizebox{18.0cm}{8.0cm}{\includegraphics{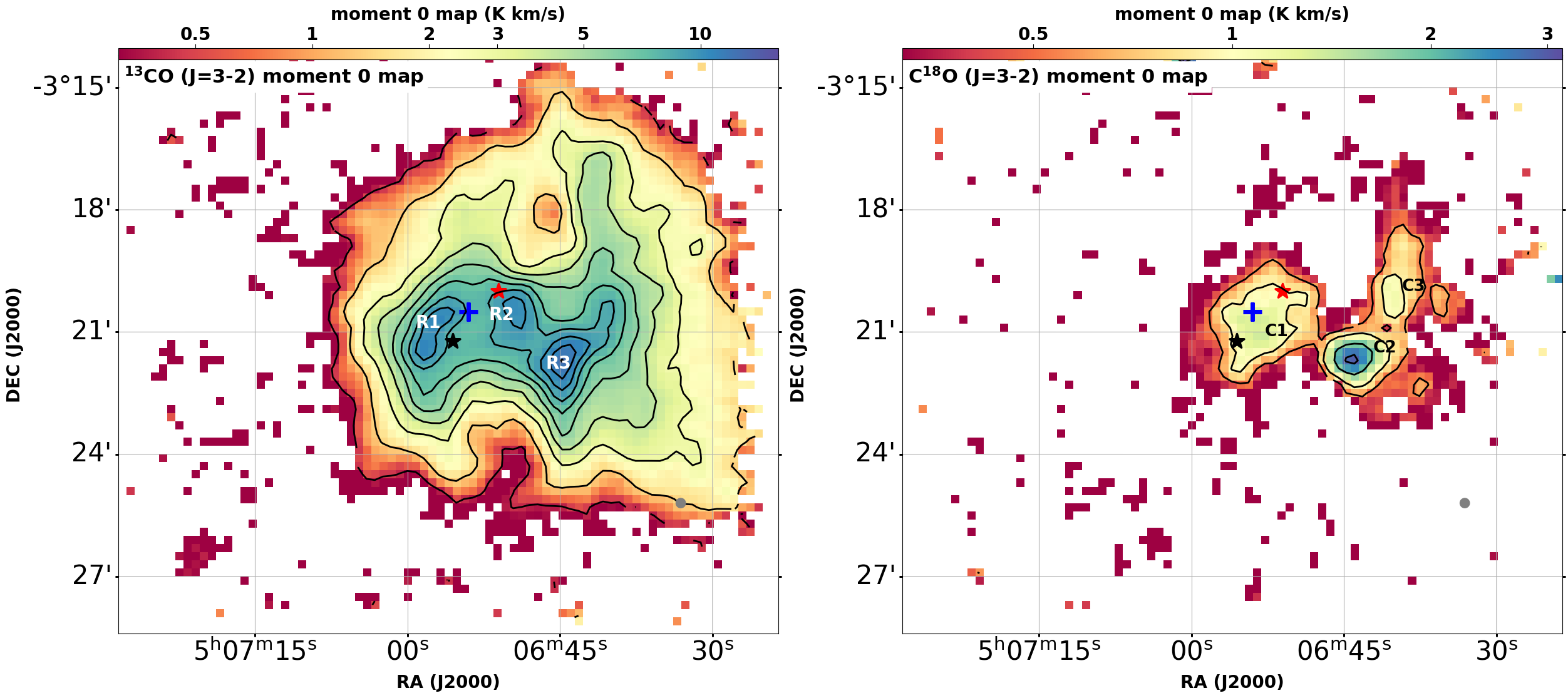}}
\caption{\textbf{Left:} The integrated intensity (moment-0) map of $^{13}\mathrm{CO}$, spanning the velocity range from 6.35 km s$^{-1}$ to 9.12 km s$^{-1}$, is illustrated for the cometary globule L1616, featuring contour levels at 0.41, 1.38, 2.76, 4.13, 5.51, 6.89, 8.27, 9.64, 11.02, and 12.40 K km s$^{-1}$. These contours are defined at 3$\sigma$ above the background emission, where $\sigma \approx 0.14$ K km s$^{-1}$ represents the standard deviation of the background noise. Within this map, three distinct clumpy regions, designated as R1, R2, and R3, have been identified as zones of particularly intense $^{13}\mathrm{CO}$ emission.
\textbf{Right:} The integrated intensity (moment-0) map of $\mathrm{C^{18}O}$, covering the same velocity range from 6.35 km s$^{-1}$ to 9.12 km s$^{-1}$, is presented for the cometary globule L1616, with contour levels set at 0.55, 0.92, 1.85, 2.77, and 3.69 K km s$^{-1}$. These contours are delineated at 3$\sigma$ above the background level, where $\sigma \approx 0.18$ K km s$^{-1}$ denotes the standard deviation of the background emission. Within this map, three prominent clumpy regions, identified as C1, C2, and C3, are recognized as areas of particularly intense $\mathrm{C^{18}O}$ emission. The blue plus sign indicates the location of the reflection nebula NGC 1788, while the black and red stars mark the positions of the two intermediate-mass stars, Kiso A-0974 15 and HD 293815, respectively.
}\label{Fig: 13CO and C18O moment 0 maps}
\end{center}
\end{figure*}

Figure~\ref{Fig: 13CO and C18O moment 0 maps} shows the integrated intensity (moment-0) maps of \(^{13}\)CO (left) and C\(^{18}\)O (right) over the velocity range 6.35--9.12~km~s\(^{-1}\), with contours at 3\(\sigma\) above the background noise (\(\sigma\) is the standard deviation of the background emission). In \(^{13}\)CO, emission is dominated by three luminous regions---R1, R2, and R3 (east to west)---which exhibit higher average integrated intensities than the surrounding areas. The mean, peak, and median integrated intensities within the \(^{13}\)CO emission region, delineated by the outermost contour at 0.41~K~km~s\(^{-1}\), are \(3.33 \pm 0.05\), \(12.49 \pm 0.05\), and \(2.54 \pm 0.05\)~K~km~s\(^{-1}\), respectively.  

The C\(^{18}\)O map, constructed over the same velocity range and contour threshold, highlights three bright clumps---C1, C2, and C3 (east to northwest)---concentrated in the denser central regions. The mean, peak, and median integrated intensities within the C\(^{18}\)O emission region, bounded by the outermost contour at 0.55~K~km~s\(^{-1}\), are \(0.98 \pm 0.07\), \(2.97 \pm 0.07\), and \(0.88 \pm 0.07\)~K~km~s\(^{-1}\), respectively. Moment-0 uncertainties were derived by propagating the rms noise across the integrated velocity channels.  

Spatial comparison shows that \(^{13}\)CO regions R1 and R2 coincide with C\(^{18}\)O clump C1, while R3 corresponds to C2. The higher optical depth of \(^{13}\)CO enhances the contrast between R1 and R2, whereas optically thin C\(^{18}\)O more effectively traces dense interiors, making C1, C2, and C3 prominent. Given its sensitivity to dense gas, the subsequent clump analysis (Section~\ref{section: Physical properties of clumps}) focuses on C1, C2, and C3 using C\(^{18}\)O data.

\begin{figure*}
\begin{center}
\resizebox{18.0cm}{13.0cm}{\includegraphics{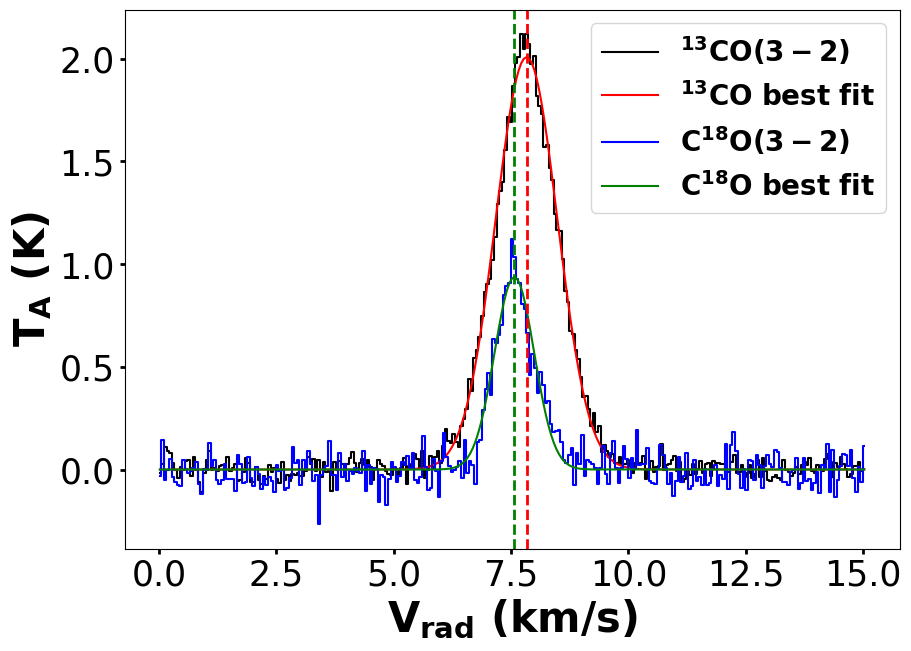}}
\caption{The average spectral profiles of \(^{13}\)CO (3-2) (in black) and C\(^{18}\)O (3-2) (in blue) emissions towards the entire \(^{13}\)CO and C\(^{18}\)O emission regions in the L1616 cometary globule. The red and green solid curves represent the Gaussian fits applied to the $^{13}$CO and C$^{18}$O spectra. The systemic velocities, indicated by the red and green dashed lines, are approximately 7.83 km s\(^{-1}\) for \(^{13}\)CO (3-2) and 7.57 km s\(^{-1}\) for C\(^{18}\)O (3-2), respectively.
}\label{Fig: 13CO and C18O spectral profiles}
\end{center}
\end{figure*}

Figure~\ref{Fig: 13CO and C18O spectral profiles} illustrates the spectral profiles of the entire \(^{13}\)CO (3-2) emission region (in black) and the C\(^{18}\)O (3-2) emission region (in blue), as presented in Figure~\ref{Fig: 13CO and C18O moment 0 maps}. The Gaussian fits over these spectra (in red for \(^{13}\)CO and green for C\(^{18}\)O) facilitate the precise determination of the systemic velocities ($V_{\rm{LSR}}$), revealing the bulk motion of gas within the L1616 PDR. The systemic velocities are estimated to be approximately 7.83 km $s^{-1}$ for \(^{13}\)CO and 7.57 km $s^{-1}$ for C\(^{18}\)O, as indicated by the red and green dashed lines on the spectral profiles, respectively. 

\begin{figure*}
\begin{center}
\resizebox{18.0cm}{8.0cm}{\includegraphics{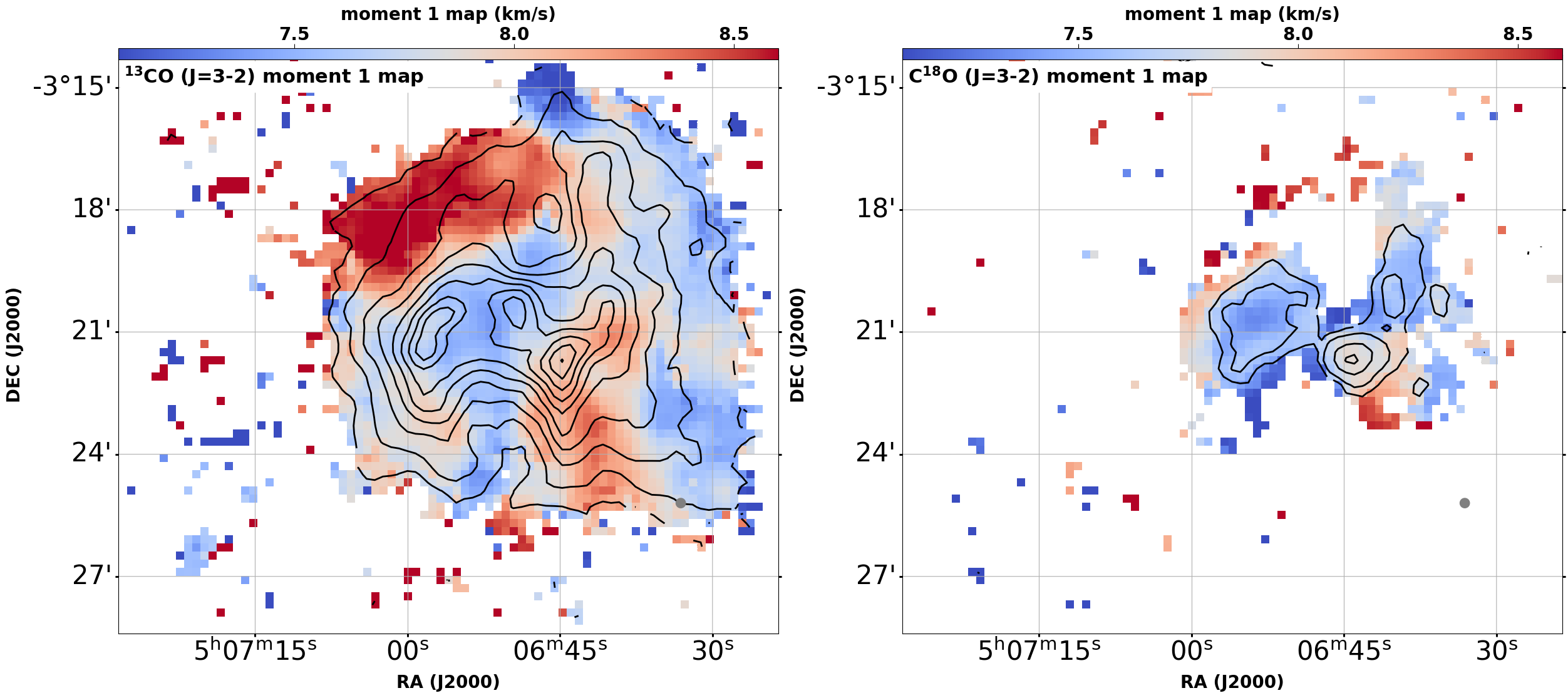}}
\caption{\textbf{Left:} $\tco$ intensity-weighted mean velocity (moment-1) map created in the velocity range between 6.35 km $s^{-1}$ and 9.12 km $s^{-1}$ overlaid with $^{13}$CO integrated intensity contours. Contour levels are at 0.41, 1.38, 2.76, 4.13, 5.51, 6.89, 8.27, 9.64, 11.02, and 12.40 K km $s^{-1}$. \textbf{Right:} C$^{18}$O intensity-weighted mean velocity (moment-1) map created in the same velocity range between 6.35 km $s^{-1}$ and 9.12 km $s^{-1}$ overlaid with C$^{18}$O integrated intensity contours. Contour levels are at 0.55, 0.92, 1.85, 2.77, and 3.69 K km $s^{-1}$.}\label{Fig: 13CO and C18O moment 1 map}
\end{center}
\end{figure*}

Figure~\ref{Fig: 13CO and C18O moment 1 map} presents the intensity-weighted mean velocity (moment-1) maps of \(^{13}\)CO (left) and C\(^{18}\)O (right) emissions, both constructed over the same velocity range as their corresponding moment-0 maps. These maps are overlaid with the integrated intensity contours of \(^{13}\text{CO}\) and C\(^{18}\)O from Figure~\ref{Fig: 13CO and C18O moment 0 maps}. In the L1616 PDR, the northeastern area and the R3  region within the \(^{13}\text{CO}\) emission display redshifted velocities, while the R1 and R2 regions exhibit blueshifted velocities relative to the systemic velocity of approximately 7.83 km s\(^{-1}\). The mean moment-1 value within the entire \(^{13}\text{CO}\) emission region is estimated to be 7.88 km s\(^{-1}\), with the velocity range extending from a maximum of approximately 8.77 km s\(^{-1}\) to a minimum of about 6.77 km s\(^{-1}\).

In the C\(^{18}\)O emission, the C2 clump region (coinciding with R3) exhibits redshifted velocities, while the C1 clump (associated with R1 and R2) and C3 clump display blueshifted velocities relative to the systemic velocity of 7.57 km s\(^{-1}\) for C\(^{18}\)O. The mean moment-1 value within the entire C\(^{18}\)O emission region is estimated to be 7.62 km s\(^{-1}\), with the velocity range extending from a maximum of approximately 8.03 km s\(^{-1}\) to a minimum of about 7.31 km s\(^{-1}\).

\begin{figure*}
\begin{center}
\resizebox{18.0cm}{8.0cm}{\includegraphics{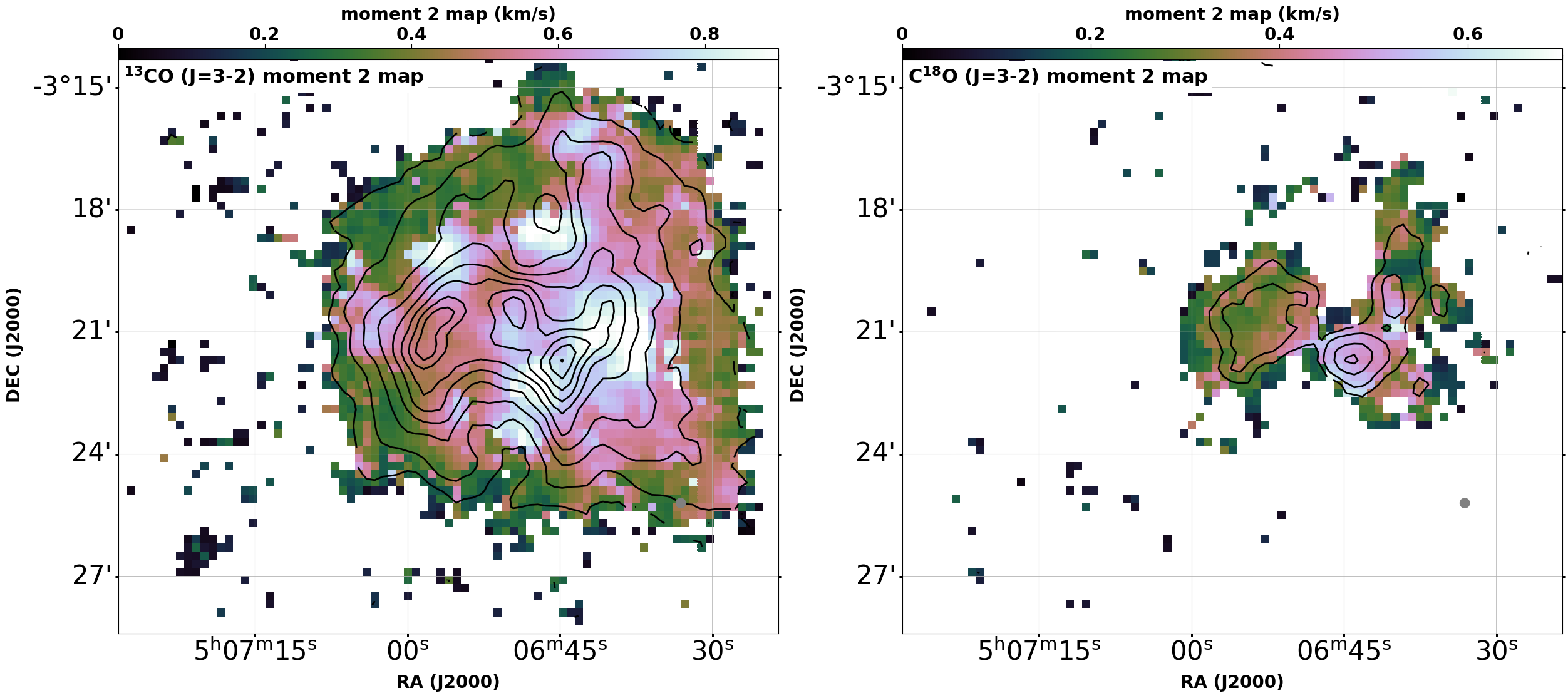}}
\caption{\textbf{Left:} $\tco$ velocity dispersion (moment-2) map created in the velocity range between 6.35 km $s^{-1}$ and 9.12 km $s^{-1}$ overlaid with $^{13}$CO integrated intensity contours. Contour levels are at 0.41, 1.38, 2.76, 4.13, 5.51, 6.89, 8.27, 9.64, 11.02, and 12.40 K km $s^{-1}$. \textbf{Right:} C$^{18}$O velocity dispersion (moment-2) map created in the same velocity range between 6.35 km $s^{-1}$ and 9.12 km $s^{-1}$ overlaid with C$^{18}$O integrated intensity contours. Contour levels are at 0.55, 0.92, 1.85, 2.77, and 3.69 K km $s^{-1}$.}\label{Fig: 13CO and C18O moment 2 maps}
\end{center}
\end{figure*}

In Figure~\ref{Fig: 13CO and C18O moment 2 maps}, we present the velocity dispersion (moment-2) maps of \(^{13}\text{CO}\) (left) and C\(^{18}\)O (right) emissions, generated over the same velocity range as the moment-0 maps. These maps are overlaid with the integrated intensity contours of \(^{13}\text{CO}\) and C\(^{18}\)O. Notably, the R3 clumpy region exhibits a higher velocity dispersion compared to the R1 and R2 regions within the \(^{13}\text{CO}\) emission. The moment-2 map for \(^{13}\text{CO}\) emission reveals a velocity dispersion range from 0.06 km s\(^{-1}\) to 1.40 km s\(^{-1}\), with a median value of approximately 0.52 km s\(^{-1}\) and an average of around 0.54 km s\(^{-1}\).

In the C$^{18}$O emission, the C2 clump shows a higher velocity dispersion than the C1 and C3 clumps. The moment-2 map for C\(^{18}\)O emission indicates a velocity dispersion range from 0.17 km s\(^{-1}\) to 0.66 km s\(^{-1}\), with a median value of approximately 0.37 km s\(^{-1}\) and an average of around 0.38 km s\(^{-1}\).

The observed enhancement in the velocity dispersion of the C2 clump of C\(^{18}\)O emission, corresponding to the R3 region in \(^{13}\text{CO}\) emission, likely signifies the presence of deeply embedded class 0 sources (MMS1 A to MMS1 D) within this clump as discussed in section ~\ref{section: Identification of clumps}. The radiative influence of these protostellar objects may be driving the observed increase in velocity dispersion, indicating their role in shaping the kinematic properties of the surrounding gas.

\section{Analysis}
\label{section: Analysis}

In this section, we explored the physical characteristics of the denser clumps, as traced by C$^{18}$O (3-2) emission. To achieve this, we first delineated the extent of the $^{13}$CO cloud and identified the distinct C$^{18}$O clumps, followed by a detailed analysis of the physical properties of the clumps.

\subsection{Identification of $^{13}$CO Cloud and C$^{18}$O Clumps}
\label{section: Identification of clumps}

The $^{13}$CO emission region within the L1616 photodissociation region is delineated by the area where the integrated intensity exceeds the 3$\sigma$ threshold. This covers a wide area where WISE 12 micron emission is emitted (see Figure~\ref{Fig: WISE two color (12 micron and 4.6 micron) image map}).

\begin{figure*}
\begin{center}
\resizebox{18.0cm}{10.0cm}{\includegraphics{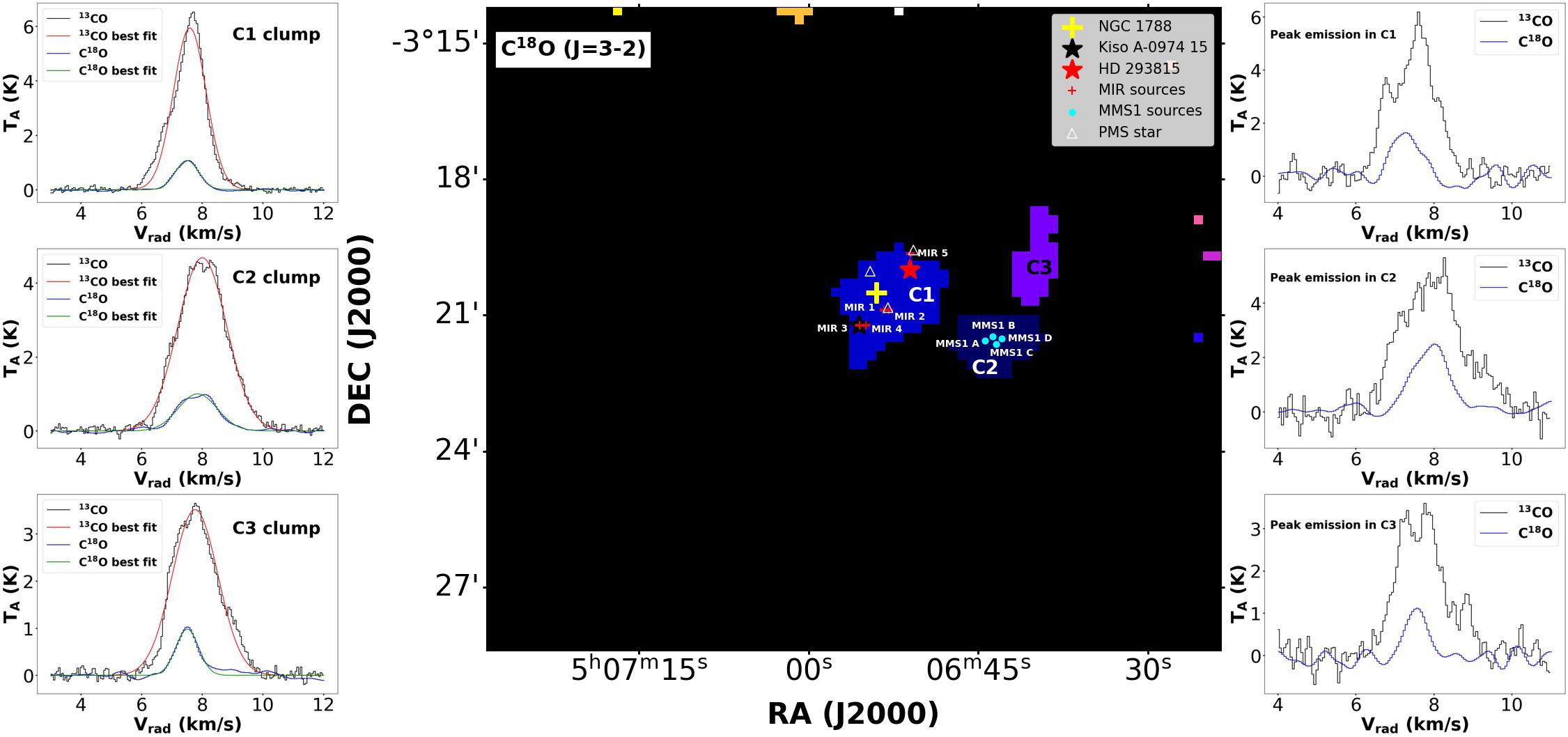}}
\caption{\textbf{Middle}: Identification of clumps using the Clumpfind algorithm. The three clumps, C1, C2, and C3, are delineated in the C$^{18}$O emission region. The yellow plus sign marks the location of the reflection nebula NGC 1788, whereas the red and black stars denote the positions of the stars HD 293815 \citep{gandolfi2008star, ramesh1995study} and Kiso A-0974 15 \citep{stanke2002triggered}, respectively. The red plus signs indicate the locations of MIR sources, and the cyan dots represent the positions of MMS1 sources \citep{stanke2002triggered}. Two pre-main sequence (PMS) stars are embedded within the C1 clump, while a third is situated in its immediate vicinity denoted by the white triangles \citep{alcala2004multi}. \textbf{Left:} Average $^{13}$CO (3--2) (black) and C$^{18}$O (3--2) (blue) spectral profiles corresponding to clumps C1 (top), C2 (middle), and C3 (bottom). The red and green solid curves represent the average Gaussian fits to the $^{13}$CO and C$^{18}$O spectra, respectively. \textbf{Right:} Spectral line profiles of $^{13}$CO (3--2) emission (black) and C$^{18}$O (3--2) emission (blue) corresponding to the clumps C1 (top), C2 (middle), and C3 (bottom), extracted towards the pixel located at the peak of the C$^{18}$O emission in each respective clump.
}\label{Fig: Clump finding using Clumpfind algorithm}
\end{center}
\end{figure*}

The identification of clumps in the C$^{18}$\text{O} emission data was rigorously confirmed using the Clumpfind algorithm, which necessitates the specification of a threshold value as an input parameter to effectively identify and delineate clumps within the emission profiles \citep{BERRY201522}. This method was adeptly applied, as depicted in the middle panel of  Figure~\ref{Fig: Clump finding using Clumpfind algorithm}, which illustrates the three clumps—C1, C2, and C3—detected in C$^{18}$\text{O} emission. The figure clearly demonstrates the spatial correspondence between clump C2 and region R3, where the peak C$^{18}$O emission is observed at RA: 5$^\text{h}$06$^\text{m}$44$^\text{s}$, DEC: -3$^\circ$21$^\prime$42$\arcsec$. It also indicates that portions of clump C1 manifest in the $^{13}$CO emission as the R1 and R2 regions with peak C$^{18}$O emission at RA: 5$^\text{h}$06$^\text{m}$56$^\text{s}$, DEC: -3$^\circ$20$^\prime$30$\arcsec$, where the optical depth of $^{13}$CO emission is relatively lower compared to the rest of the C1 clump. However, due to the higher optical depth, clump C3 with peak C$^{18}$O emission at position RA: 5$^\text{h}$06$^\text{m}$38$^\text{s}$, DEC: -3$^\circ$19$^\prime$30$\arcsec$, does not exhibit a corresponding clumpy region in the $^{13}$CO emission. Since the optical depth of clump C3 ($\bar\tau_{18}(\text{C3}) \approx 0.39$) exceeds that of the clumps C1 ($\bar\tau_{18}(\text{C1}) \approx 0.24$) and C2 ($\bar\tau_{18}(\text{C2}) \approx 0.29$) in C$^{18}$O emission, as derived in section~\ref{section: Physical properties of clumps}, it is reasonable to expect a correspondingly higher optical depth in $^{13}$CO emission, which could potentially account for its non-detection. This figure also illustrates that clump C1 encompasses the location of the reflection nebula NGC 1788, which is illuminated by a cluster of stars, including two of the brightest intermediate-mass stars, HD 293815 \citep{ramesh1995study} and Kiso A-0974 15, marked by red and black stars, respectively. Additionally, clump C1 encompasses the positions of mid-infrared sources (MIR 1-5) at $11.9 \mu$m, as identified by \citet{stanke2002triggered} taken from the 2MASS survey, where they are marked by red plus signs. It is noteworthy that the positions of the MIR 3 and MIR 4 sources align with the location of the star Kiso A-0974 15, a phenomenon likely attributable to the observational resolution. This clump harbors three pre-main sequence (PMS) stars, depicted as white triangles, as identified in the study by \cite{alcala2004multi}. Among them, the stars located at RA: 5$^\text{h}$06$^\text{m}$51$^\text{s}$, DEC: -3$^\circ$19$^\prime$34$\arcsec$, and RA: 5$^\text{h}$06$^\text{m}$53$^\text{s}$, DEC: -3$^\circ$20$^\prime$50$\arcsec$ correspond to L1616 MIR5 and L1616 MIR1, L1616 MIR2, respectively. The third star, positioned at RA: 5$^\text{h}$06$^\text{m}$55$^\text{s}$, DEC: -3$^\circ$20$^\prime$02$\arcsec$, has been classified as Kiso A-0974-14 in their study and is also recognized in \cite{stanke2002triggered}. Notably, the positions of the MIR sources identified by \cite{alcala2004multi} exhibit slight offsets from those reported by \cite{stanke2002triggered}. Meanwhile, clump C2 surrounds the 1.2 mm dust continuum sources, specifically MMS1, depicted by cyan dots. The MMS1 source is further resolved into several subcomponents, with MMS1 A being the most luminous. Other fainter sources include MMS1 B, located 10" west and 6" north; MMS1 C, positioned 15" west and 4" south; and MMS1 D, situated 22" west and 3" north relative to MMS1 A, as described by \citet{stanke2002triggered}.

\subsection{Physical properties of C$^{18}$O clumps}
\label{section: Physical properties of clumps}

The pixel count and corresponding coordinates for each clump region were initially determined, followed by the extraction of individual pixel spectra. Subsequently, the average spectrum for each clump was derived by aggregating the spectra from all pixels within the region.

The left panel of Figure~\ref{Fig: Clump finding using Clumpfind algorithm} presents the average spectral profiles of the three clumps—C1 (top), C2 (middle), and C3 (bottom)—in $^{13}$CO (black) and C$^{18}$O (blue) emission. While the averaging of spectra across all pixels within each clump tends to smooth out localized features, the self-absorption signature typically observed in $^{13}$CO emission within dense regions is not clearly evident in clump C1, likely due to this averaging effect. However, in clumps C2 and C3, the $^{13}$CO profiles exhibit faint indications of self-absorption, suggesting the presence of optically thick gas components. The velocity of the Local Standard of Rest ($V_{\text{LSR}}$) reflecting the bulk motion of the clump regions, corresponding to the velocity at
which the majority of gas molecules move, derived from Gaussian fits to the average C$^{18}$O spectra of clumps C1, C2, and C3 are $7.50 \pm 0.01$ km s$^{-1}$, $7.85 \pm 0.01$ km s$^{-1}$, and $7.51 \pm 0.01$ km s$^{-1}$, respectively. The quoted uncertainties correspond to the standard errors obtained from the covariance matrices of the Gaussian fits. These measurements align well with the observations presented at the moment 1 map of C$^{18}$O emission (Figure~\ref{Fig: 13CO and C18O moment 1 map}). 

The one-dimensional velocity dispersion ($\sigma_{1D}$) values derived from the Gaussian fits over the  C$^{18}$O spectral profiles of the clumps are $0.35 \pm 0.01$\,km\,s$^{-1}$ for C1, $0.62 \pm 0.01$\,km\,s$^{-1}$ for C2, and $0.32 \pm 0.01$\,km\,s$^{-1}$ for C3, which are consistent with the distributions observed in the moment 2 map (Figure ~\ref{Fig: 13CO and C18O moment 2 maps}). The uncertainties in velocity dispersion represent the standard errors derived from the covariance matrices associated with the Gaussian profile fittings. The relatively enhanced velocity dispersion in clump C2 is attributed to protostellar outflow activity associated with the embedded Class 0 protostar system MMS1 A--D \citep{stanke2002triggered}. In contrast, clump C1, which hosts a population of PMS stars \citep{stanke2002triggered, alcala2004multi, gandolfi2008star, ramesh1995study}, shows no signatures of active outflow, resulting in a comparatively lower velocity dispersion. Clump C3, devoid of any evident star formation activity, exhibits the narrowest velocity dispersion, consistent with a quiescent molecular environment. 

Astronomical observations targeting a specific molecular transition are most responsive to gas densities that approximate the critical density of the tracer \citep{stahler2008formation}. Under the assumption of local thermodynamic equilibrium (LTE), the gas density is considered to be equal to this critical density. Accordingly, all physical parameters of the clump structures have been derived under LTE conditions, wherein the population distribution of molecular energy levels adheres to the Boltzmann statistics. Within the LTE framework, the kinetic temperature of the gas is presumed to be equivalent to the excitation temperature.

Due to the unavailability of $^{12}$CO (3--2) observations for the L1616 cloud, the excitation temperatures of the clumps have been estimated by assuming that the $^{13}$CO (3--2) transition is optically thick within the identified dense clump regions. This assumption is physically justified, as the \(^{12}\)CO line is often subject to significant self-absorption in high-density environments, which can render it unsuitable for reliable temperature estimation. In such regions, \(^{13}\)CO becomes optically thick and can thus serve as a more robust tracer of excitation conditions \citep{myers1983dense, buckle2010jcmt}. This is further supported by the presence of self-absorption features evident in the $^{13}$CO (3--2) spectral profiles (shown in black) in the right panel of Figure~\ref{Fig: Clump finding using Clumpfind algorithm}, which correspond to the pixels at the peak of C$^{18}$O emission in each clump. Similar self-absorption signatures are also observed in other pixels within the clumps, providing strong evidence that the $^{13}$CO emission is indeed optically thick in these regions. Accordingly, the use of \(^{13}\)CO emission to derive excitation temperatures is well-supported by previous studies \citep{buckle2010jcmt} and is particularly appropriate in the context of dense clump analysis. Under this assumption and within the LTE approximation, it is further presumed that the relative populations in the lower rotational states of the $^{13}$CO and C$^{18}$O isotopologues are identical. Consequently, the excitation temperatures derived from $^{13}$CO (3 - 2) emission are assumed to be representative of those for the C$^{18}$O (3 - 2) transition as well.

The excitation temperature of the molecular gas can be estimated by employing the standard radiative transfer equation for an isothermal slab. This formulation, commonly known as the detection equation, quantifies the observed emission line intensity as a function of the excitation temperature, background radiation, and optical depth \citep{mangum2015calculate, rawat2024giant, stahler2008formation}:
\begin{equation}
    T_{mb}^{peak} = f\left[J\left(T_{ex}\right) - J\left(T_{BG}\right)\right] \left[1 - \exp\left(-\tau_{13}\right)\right],
    \label{detection equation}
\end{equation}
where $\tau_{13}$ denotes the optical depth at the line center of the $^{13}$CO spectral profile, and $T_{mb}^{peak}$ is the peak main-beam temperature. The main-beam temperature is related to the observed antenna temperature through:

\begin{equation}
    T_{mb}^{peak} = \frac{T_{A}^{peak}}{\eta_{mb}},
    \label{main beam equation}
\end{equation}
where $\eta_{mb} = 0.61$ represents the main beam efficiency for both the $^{13}$CO (3-2) and C$^{18}$O (3--2) transitions, adopted from \citet{buckle2010jcmt}. The beam filling factor $f$ is assumed to be unity \citep{buckle2010jcmt, rawat2024giant}.

The function $J(T)$ is given by:

\begin{equation}
    J(T) = \frac{T_{0}}{\exp\left(\frac{T_0}{T}\right) - 1},
    \label{J(T) equation}
\end{equation}
where $T_0$ is defined as:

\begin{equation}
    T_0 = \frac{h\nu}{k_B},
    \label{T0 equation}
\end{equation}
with \( h \) being Planck’s constant, \( \nu \) the rest frequency of the molecular transition, and \( k_B \) the Boltzmann constant. For the transitions analyzed in this study, \( T_0 \) takes on values of 15.89\,K for the $^{13}$CO (3--2) line and 15.81\,K for the C$^{18}$O (3--2) line, respectively, based on their rest frequencies. The background radiation temperature, \( T_{\mathrm{BG}} \), is assumed to be 2.7\,K, corresponding to the cosmic microwave background (CMB), and is accounted for in radiative transfer calculations as a baseline against which line excitation is measured.

Under the assumption that the $^{13}$CO (3--2) transition is optically thick, the second term in Equation \ref{detection equation} asymptotically approaches unity, thereby simplifying the expression. Substituting Equation~\ref{J(T) equation} and adopting the corresponding values of \( T_0 \) and \( T_{BG} \), the excitation temperature can be expressed analytically as:

\begin{equation}
    T_{ex}^{13} (3 \rightarrow 2) = \frac{15.89\,\mathrm{K}}{\ln\left[1 + \frac{15.89\,\mathrm{K}}{T_{mb, peak}^{13} + 0.044\,\mathrm{K}}\right]}.
    \label{excitation temperature equation}
\end{equation}
Using Equation~\ref{excitation temperature equation}, the excitation temperature (\( T_{\text{ex}} \)) was computed on a pixel-by-pixel basis within each identified clump. For the C1 clump, \( T_{\text{ex}} \) ranges from a maximum of $24.6 \pm 0.2$ K to a minimum of $13.1 \pm 0.2$ K, with a mean value of approximately 16.7\,K and a median of 16.6\,K. In the C2 clump, the excitation temperature spans from $19.4 \pm 0.2$ K to $10.6 \pm 0.1$ K, yielding a mean and median of 14.3\,K and 14.1\,K, respectively. The C3 clump exhibits the lowest excitation temperatures, varying between $13.7 \pm 0.2$ K and $10.8 \pm 0.2$ K, with corresponding mean and median values of 12.3\,K and 12.4\,K. The uncertainties in excitation temperature are propagated from the uncertainty in the antenna temperature ($T_A$), which was derived from the standard errors of the Gaussian amplitude fitted to the spectral line profiles.

These spatial variations in excitation temperature reflect the distinct star formation activity within each clump. The elevated temperatures in C1 are attributable to the presence of PMS stars, which contribute additional thermal energy to the gas. Although C2 harbors Class 0 protostars, its excitation temperature remains lower than that of C1, likely due to the deeply embedded nature and early evolutionary stage of these protostars. In contrast, C3 shows the lowest excitation temperatures, consistent with the absence of any apparent star formation activity.

After deriving the excitation temperature, the optical depth for C$^{18}$O (3-2) emission can be derived from equation \ref{detection equation} by substituting the values of excitation temperature. The optical depth equation can be written as follows,

\begin{equation}
    \tau_{0}^{18} = -\ln\left(1 - \frac{T^{18}_{{mb, peak}}}{15.81}\left[\frac{1}{\exp\left(\frac{15.81}{T_\text{ex}}\right) - 1} - \left(2.89 \times 10^{-3}\right)\right]^{-1}\right).
    \label{eq: c18o tau equation}
\end{equation}
Using Equation~\ref{eq: c18o tau equation}, the optical depth of the C$^{18}$O (3--2) emission line was calculated on a pixel-by-pixel basis within each clump. In the C1 clump, the optical depth ranges from a minimum of 0.03 to a maximum of 0.46, with an average value of approximately 0.24. For the C2 clump, the optical depth varies between 0.08 and 0.73, yielding a mean value of about 0.29. In contrast, the C3 clump exhibits the highest optical depths, spanning from 0.13 to 0.75, with a mean around 0.39.

Additional physical parameters, such as the line width of spectral profiles and the three-dimensional velocity dispersion, are derived using the following two fundamental equations:
\begin{equation}
    \Delta V = 2.35\sigma_{1D},
    \label{line width equation}
\end{equation}
\begin{equation}
    \sigma_{3D} = \sqrt{3} \sigma_{1D}.
    \label{3d velocity dispersion equation}
\end{equation}
The three-dimensional velocity dispersion provides a more holistic view of the internal motions within the cloud, considering the full velocity distribution across all spatial dimensions. Together, these parameters offer a comprehensive characterization of the molecular cloud’s physical and dynamical properties, contributing to our understanding of its structure, composition, and evolutionary processes.

Table \ref{physical properties clumps} provides a detailed summary of the key physical parameters characterizing the clumps C1, C2, and C3 within the L1616 photodissociation region. The listed quantities include the average excitation temperature (T$_\mathrm{ex}$), optical depth ($\tau$), full width at half maximum (FWHM or $\Delta V$), three-dimensional velocity dispersion ($\sigma_\mathrm{3D}$), and the systemic velocity (V$_\mathrm{LSR}$), offering critical insights into the local kinematics and thermodynamic conditions of each clump.

\subsubsection{Column density, mass, and virial parameters estimation}
The physical extent of each clump was determined by calculating the total area based on the pixel count and pixel scale, using the following relation:
\begin{equation}
    r_{\mathrm{eff}} = \sqrt{\frac{A_{\mathrm{eff}}}{\pi}},
    \label{effective radii equation}
\end{equation}
where \( A_{\mathrm{eff}} \) denotes the effective projected area of a clump. Adopting a distance of 384 pc to the L1616 cometary globule \citep{saha2022magnetic}, the resulting effective radii for the clumps C1, C2, and C3 are estimated to be approximately $0.128 \pm 0.002$ pc, $0.093 \pm 0.001$ pc, and $0.078 \pm 0.001$ pc, respectively. The uncertainty in the determination of $r_{\mathrm{eff}}$ primarily arises from the uncertainty associated with the distance measurement.

Utilizing the derived average excitation temperature values, as discussed above, we have successfully computed the average column densities pertaining to the C1, C2, and C3 clump  regions within the L1616 PDR using the following relations \citep{garden1991spectroscopic, buckle2010jcmt},
\begin{equation}
    N(\mathrm{C^{18}O}) = 8.26\times 10^{13} \exp\left(\frac{15.81}{T_{ex}}\right)\times \frac{T_{ex}+0.88}{1-\exp\left(\frac{-15.81}{T_{ex}}\right)}\int \tau \,dv \,\mathrm{cm^{-2}},
    \label{column density equation}
\end{equation}
where we have used the approximations,
\begin{equation}
    \int \tau \, dv = \frac{1}{J({T_{ex}}) - J({T_{BG}})} \frac{\tau}{1 - e^{-\tau}} \int T_{mb} \, dv \left( \text{for } \tau \geq 1 \right),
    \label{optical depth correction equation}
\end{equation}
and
\begin{equation}
    \int \tau \, dv = \frac{1}{J({T_{ex}}) - J({T_{BG}})} \int T_{mb} \, dv \left( \text{for } \tau \ll 1 \right).
    \label{no optical depth equation}
\end{equation}

The hydrogen column density based on C$^{18}$O (3-2) \citep{castets1995physical, rawat2024giant} emission are obtained
using the following equation,
\begin{equation}
    (N(H_2))_{\mathrm{C^{18}O}} = 7 \times 10^6 N(\mathrm{C^{18}O}).
    \label{hydrogen column density equation}
\end{equation}
Using equations \ref{column density equation}, \ref{no optical depth equation}, and \ref{hydrogen column density equation}, the column densities of C$^{18}$O and H$_{2}$ molecules were derived for each pixel within the identified clumps without optical depth correction because C$^{18}$O (3-2) emission is optically thin in the clumps.

For clump C1, the C$^{18}$O column density spans from $1.3 \times 10^{14}$ cm$^{-2}$ to $2.2 \times 10^{15}$ cm$^{-2}$, corresponding to an H$_{2}$ column density range of $9.0 \times 10^{20}$ cm$^{-2}$ to $1.5 \times 10^{22}$ cm$^{-2}$. The mean column densities across C1 are calculated to be $1.0 \times 10^{15}$ cm$^{-2}$ for C$^{18}$O and $(7.2 \pm 1.2) \times 10^{21}$ cm$^{-2}$ for H$_{2}$.

In the case of clump C2, the C$^{18}$O column density ranges from $5.6 \times 10^{14}$ cm$^{-2}$ to $4.1 \times 10^{15}$ cm$^{-2}$, with the associated H$_{2}$ column density varying between $3.9 \times 10^{21}$ cm$^{-2}$ and $2.9 \times 10^{22}$ cm$^{-2}$. The average values within this clump are $1.8 \times 10^{15}$ cm$^{-2}$ for C$^{18}$O and $(1.3 \pm 0.2) \times 10^{22}$ cm$^{-2}$ for H$_{2}$.

For clump C3, the C$^{18}$O column density lies between $8.4 \times 10^{14}$ cm$^{-2}$ and $2.2 \times 10^{15}$ cm$^{-2}$, while the corresponding H$_{2}$ column density ranges from $5.9 \times 10^{21}$ cm$^{-2}$ to $1.6 \times 10^{22}$ cm$^{-2}$. The mean column densities within C3 are estimated to be $1.5 \times 10^{15}$ cm$^{-2}$ for C$^{18}$O and $(1.1 \pm 0.1) \times 10^{22}$ cm$^{-2}$ for H$_{2}$. The uncertainty in the column density arises through the propagation of errors, primarily stemming from the uncertainty in the excitation temperature, which plays a crucial role in its derivation.

The molecular gas mass within each clump has been calculated using the following relation:
\begin{equation}
    M_{c} = \mu_{H_{2}}m_{H}A_{pixel}\sum N(H_{2}),
    \label{mass equation}
\end{equation}
Here, $\mu_{H_2}$ represents the mean molecular weight, adopted as 2.8 \citep{kauffmann2008mambo}, $m_{H}$ denotes the mass of a hydrogen atom, and $A_{\text{pixel}}$ is the area of a pixel in units of cm$^{2}$. The resulting molecular gas masses for the clumps C1, C2, and C3 are estimated to be $8.2 \pm 0.2$~M$_\odot$, $7.7 \pm 0.2$~M$_\odot$, and $4.5 \pm 0.1$~M$_\odot$, respectively. The uncertainty in the mass estimation is primarily propagated from the uncertainties in both the distance to the L1616 cloud and the column density.

In the study by \cite{yonekura1999search}, a single core was identified in the C$^{18}$O (1--0) emission, and its mass was estimated to be 161~M$_\odot$, based on an assumed distance of 460~pc to the L1616 cloud. Their observations were conducted with an angular resolution of 2.7$^\prime$, which is significantly coarser than the resolution of our JCMT-HARP data, approximately 15$\arcsec$. Adopting the same distance of 460~pc, we estimated the masses of the three clumps identified in our analysis to be 11.8~M$_\odot$, 11.1~M$_\odot$, and 6.5~M$_\odot$, respectively, yielding a total mass of about 29.4~M$_\odot$. The discrepancy in the mass estimates is likely attributable to the lower spatial resolution of their observations, which may have resulted in the blending of multiple structures into a single core. In contrast, our higher-resolution data reveal the presence of three distinct clumps within the L1616 cloud region.

The additional physical characteristics of the molecular clumps—particularly the volume number density of molecular hydrogen (n$_{\mathrm{H}_2}$)—were derived using the following relation \citep{rawat2023probing, pineda2008co}:
\begin{equation}
    n_{\mathrm{H}_2} = \frac{3M_{c}}{4\pi r_{\mathrm{eff}}^{3} \mu_{\mathrm{H}_2} m_{\mathrm{H}}},
    \label{number density equation}
\end{equation}

The derived average values of number density for the clumps C1, C2, and C3 are $(1.4 \pm 0.6)$ $\times$ 10$^{4}$ cm$^{-3}$, $(3.3 \pm 1.5)$ $\times$ 10$^{4}$ cm$^{-3}$, and $(3.3 \pm 1.6)$ $\times$ 10$^{4}$ cm$^{-3}$, respectively. The uncertainties in the derived number densities originate from the propagated errors in both the mass and effective radius of the clumps.

To evaluate the gravitational boundedness and dynamical stability of the clumps, we computed the virial mass (M$_{\mathrm{vir}}$) and the virial parameter ($\alpha$), assuming spherical geometry and a radial density profile of the form $\rho \propto r_{\mathrm{eff}}^{-\beta}$ with $\beta = 2$ \citep{rawat2023probing}, using the following equations,
\begin{equation}
    M_{\mathrm{vir}} = 126\, r_{\mathrm{eff}} \left(\Delta V\right)^{2},
    \label{virial mass}
\end{equation}
\begin{equation}
    \alpha = \frac{M_{\mathrm{vir}}}{M_{c}}.
    \label{virial parameter}
\end{equation}
The virial masses estimated for the C1, C2, and C3 clumps are $10.8 \pm 1.6$ M$_\odot$, $20.9 \pm 1.5$ M$_\odot$, and $8.9 \pm 3.2$ M$_\odot$, respectively. The uncertainty in the virial mass arises from the propagated errors in the clump radius and the velocity dispersion, the latter derived from the standard deviation of the Gaussian fit to the C$^{18}$O spectral line profile.
Notably, the FWHM, $\Delta V$ used in Equation~\ref{virial mass} represents the average of all pixel-wise line widths within each clump, rather than the line width derived from the mean spectrum shown in the left panel of Figure~\ref{Fig: Clump finding using Clumpfind algorithm}. This method avoids potential overestimation of $\Delta V$ and yields more accurate virial mass values. The resulting virial parameters for C1, C2, and C3 are $1.3 \pm 0.2$, $2.7 \pm 0.2$, and $2.0 \pm 0.7$, respectively. The estimated uncertainties in the virial parameter arise from the propagated uncertainties in both the virial mass and the clump mass. Since the virial parameter for clump C1 falls below the critical threshold defined by the density index $\beta = 2$, this indicates that this clump is gravitationally bound. Such conditions are consistent with regions undergoing active star formation or poised on the brink of collapse. In contrast, clump C2 exhibits a virial parameter exceeding this threshold, suggesting that it is gravitationally unbound. This dynamical state may be attributed to the influence of protostellar feedback, particularly the outflows associated with the embedded Class 0 protostars, which can inject sufficient kinetic energy to counteract gravitational confinement. For clump C3, the virial parameter is found to be close to the critical value corresponding to a density profile index of 2. This suggests that C3 is likely in a state of near virial equilibrium, rather than being gravitationally bound.

A detailed summary of these derived physical parameters for clumps C1, C2, and C3 is presented in Table \ref{Physical properties calculation for clumps}.

\subsubsection{Turbulent kinematics analysis in C$^{18}$O emission in clump regions}

To explore the turbulent characteristics within the clump regions C1, C2, and C3 of the L1616 cloud, we computed the non-thermal velocity dispersion and the Mach number, following the methodologies of \cite{sepulveda2020vla} and \cite{palau2015gravity}. The three-dimensional non-thermal velocity dispersion ($\sigma_{\mathrm{nt,\,3D}}$) is related to its one-dimensional counterpart ($\sigma_{\mathrm{nt,\,1D}}$) through the expression:
\begin{equation}
    \sigma_{\mathrm{nt,\,3D}} = \sqrt{3} \, \sigma_{\mathrm{nt,\,1D}}.
    \label{eq:3D_nt_dispersion}
\end{equation}
The one-dimensional non-thermal velocity dispersion is derived from the observed total velocity dispersion ($\sigma_{\mathrm{1D}}$) and the thermal velocity dispersion ($\sigma_{\mathrm{th}}$), where the latter is given by:
\begin{equation}
    \sigma_{\mathrm{th}} = \sqrt{ \frac{k_B T_{\mathrm{kin}}}{\mu_i m_{\mathrm{H}}} },
    \label{eq:thermal_dispersion}
\end{equation}
Here, $k_B$ is the Boltzmann constant, $T_{\mathrm{kin}}$ is the kinetic temperature within each of the clump regions, and $\mu_i$ is the mean molecular weight of the tracer species (taken to be 30 for C$^{18}$O; \citep{rawat2024giant}). The non-thermal component is then calculated as:
\begin{equation}
    \sigma_{\mathrm{nt,\,1D}} = \sqrt{ \sigma_{\mathrm{1D}}^2 - \sigma_{\mathrm{th}}^2 }.
    \label{eq:nt_dispersion}
\end{equation}
To characterize the level of turbulence relative to thermal motion, we estimated the Mach number ($\mathcal{M}$), defined as the ratio of the 3D non-thermal velocity dispersion to the thermal sound speed ($c_s$):
\begin{equation}
    \mathcal{M} = \frac{ \sigma_{\mathrm{nt,\,3D}} }{ c_s }.
    \label{eq:mach_number}
\end{equation}
The sound speed, $c_s$, is given by:
\begin{equation}
    c_s = \sqrt{ \frac{k_B T_{\mathrm{kin}}}{\mu m_{\mathrm{H}}} },
    \label{eq:sound_speed}
\end{equation}
where $\mu$ represents the mean molecular weight per free particle, adopted as 2.37 for a molecular gas with typical interstellar composition \citep{kauffmann2008mambo, rawat2024giant}.

Using Equation~\ref{eq:mach_number}, we derived the Mach numbers for the C1, C2, and C3 clumps as 2.4, 4.4, and 3.4, respectively. The notably elevated Mach number observed in the C2 clump likely reflects enhanced turbulent motions, potentially driven by the outflow activity associated with embedded Class 0 protostellar sources. It is important to emphasize that the turbulent parameters presented in this analysis represent the mean of these values computed across all individual pixels within each clump. This approach mitigates the risk of overestimating turbulence parameters that can arise when using the line width obtained from a Gaussian fit to the averaged spectrum of the clumps (as shown in the left panel of Figure ~\ref{Fig: Clump finding using Clumpfind algorithm}).

Table \ref{physical properties clumps} presents the parameters that provide further insight into the turbulence levels within these clump regions.

\subsubsection{Gravitational potential energy and kinetic energy estimation}

The gravitational potential energy ($W$) of the clump regions was estimated by assuming a radial density profile of the form $\rho \propto r^{-\beta}$, in accordance with the prescriptions of \cite{maclaren1988corrections} and \cite{bertoldi1992pressure}. For a power-law index $\beta = 2$ and under the assumption of spherical symmetry, the gravitational potential energy is given by:
\begin{equation}
    W = -\frac{3}{5} \gamma \frac{G M_{c}^2}{r_{\mathrm{eff}}},
    \label{eq:potential_energy}
\end{equation}
where $\gamma = \frac{1 - \beta/3}{1 - 2\beta/5} = \frac{5}{3}$ for $\beta = 2$, and $G$ is the gravitational constant. 

The internal kinetic energy, which reflects the contribution from thermal and non-thermal turbulent motions within the gas, was evaluated using the expression \citep{buckle2010jcmt}:
\begin{equation}
    E_{\mathrm{kin}} = \frac{1}{2} M_{c} \sigma^2_{\mathrm{3D}},
    \label{eq:kinetic_energy}
\end{equation}
where $\sigma_{\mathrm{3D}}$ denotes the three-dimensional velocity dispersion averaged over all pixels within each clump.

The gravitational potential energy and kinetic energy of the clump C1 are estimated to be $-(4.6 \pm 0.3) \times 10^{36}$~J and $(3.0 \pm 0.6) \times 10^{36}$~J, respectively. For C2, these values are $-(5.6 \pm 0.3) \times 10^{36}$~J and $(7.5 \pm 0.6) \times 10^{36}$~J, while for C3, they are $-(2.3 \pm 0.1) \times 10^{36}$~J and $(2.2 \pm 0.9) \times 10^{36}$~J. 

The uncertainties associated with the gravitational energies have been propagated from the errors in clump mass and effective radius, whereas the uncertainties in kinetic energy were derived based on the propagated errors in mass and the one-dimensional velocity dispersion. 

These calculated values of gravitational binding and internal kinetic energies provide crucial insights into the dynamical states of the clumps identified in the C$^{18}$O emission. A detailed summary of these energetics is presented in Table~\ref{Physical properties calculation for clumps}.

For the C1 clump, the absolute value of the gravitational potential energy exceeds the internal kinetic energy, indicating that this structure is gravitationally bound within the C$^{18}$O emission region. In contrast, the C2 clump exhibits a kinetic energy that surpasses its gravitational potential energy, suggesting that it is gravitationally unbound and dynamically supported against collapse. For clump C3, the magnitudes of gravitational and kinetic energies are nearly equal, indicating that the clump is in a state close to virial equilibrium. This is supported by the energy ratio \( 2E_{\mathrm{kin}}/|W| \approx 2.0 \), which is consistent with the condition expected for a system in virial balance.

\begin{table*}
\begin{center}
	\caption{Kinematic and Turbulent Properties of Clumps C1, C2, and C3 in the L1616 PDR Traced by C$^{18}$O (3–2) Emission.}
	\label{physical properties clumps}
    \renewcommand{\arraystretch}{1.3} 
	\begin{tabular}{lccccccccc} 
		\hline
		Region  & $\langle T_{kin}\rangle$ & $\bar\tau$  &  $\Delta V$  & $\sigma_{3D}$ & $V_{LSR}$ & $\sigma_{th}$ & $\sigma_{nt, 3D}$ & $c_{s}$ & $\mathbf{\mathcal{M}}$ \\ 
		  & (K) & & (km $s^{-1}$) & (km $s^{-1}$) & (km $s^{-1}$) & (km $s^{-1}$) & (km $s^{-1}$) & (km $s^{-1}$) & \\ 
		\hline
    C1 clump & $16.7$ & $0.24$ & 0.83 & 0.61 & 7.50 & $0.07$ & $0.59$ & $0.24$ & $2.4$\\
    C2 clump & $14.3$ & $0.29$ & 1.46 & 1.08 & 7.85 & $0.06$ & $0.98$ & $0.22$ & $4.4$\\
    C3 clump & $12.3$ & $0.39$ & 0.76 & 0.56 & 7.51 & $0.06$ & $0.69$ & $0.21$ & $3.4$\\
		\hline 
	\end{tabular}
\end{center}
\end{table*}

\begin{table*}
\begin{center}
	\caption{Derived Physical Parameters of Clumps C1, C2, and C3 in the L1616 PDR from C$^{18}$O (3–2) Emission.}
	\label{Physical properties calculation for clumps}
        \renewcommand{\arraystretch}{1.3} 
	\begin{tabular}{lcccccccc} 
		\hline
		Region & $\langle N(H_2)\rangle$  & $r_{eff}$ & Mass  & $\langle n(H_2)\rangle$ & $M_{vir}$ & $\alpha$ & -W & $E_{kin}$\\ 
		     &($\times$$10^{22}$$cm^{-2}$) & (pc) &  ($M_{\odot}$) & ($\times$$10^4$$cm^{-3}$) & ($M_{\odot}$)  &  & ($\times10^{36}$ J) & ($\times10^{36}$ J)\\ 
		\hline
    C1 clump & $0.72$ & 0.13 & $8.2$ & $1.37$ & $10.8$ & $1.32$ & $4.58$ & $3.00$\\
    C2 clump & $1.27$ & 0.09 & $7.7$ & $3.29$ & $20.9$ & $2.70$ & $5.55$ & $7.46$\\
    C3 clump & $1.07$ & 0.08 & $4.5$ & $3.35$ & $8.9$ & $1.96$ & $2.27$ & $2.23$\\
		\hline 
	\end{tabular}
\end{center}
\end{table*}

\subsection{Gas Kinematics of $^{13}$CO Cloud and C$^{18}$O Clumps}

\begin{figure*}
\begin{center}
\resizebox{13.0cm}{11.5cm}{\includegraphics{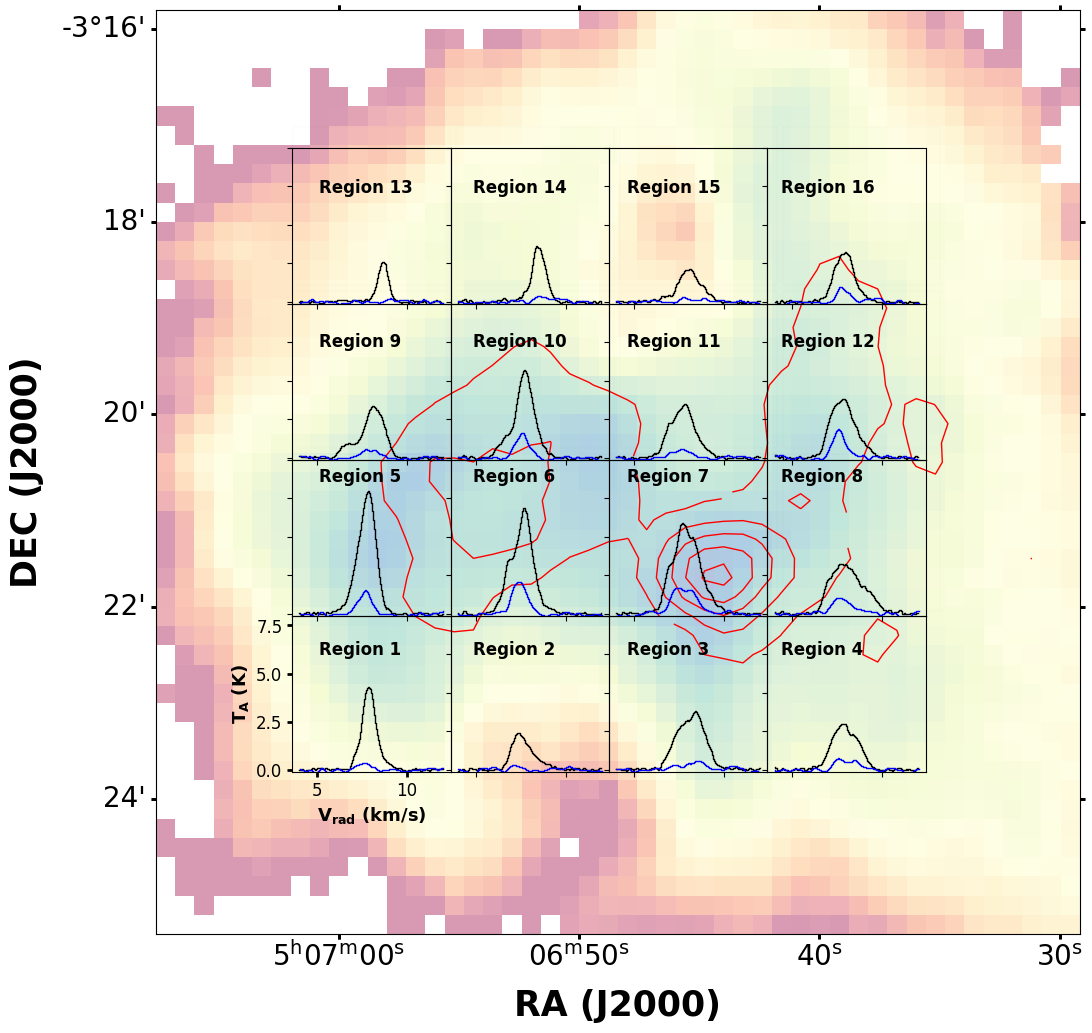}}
\caption{$^{13}\text{CO}$ (3-2) (in black) and C$^{18}$O (in blue) spectral profiles of 16 regions, each with a size of 98\arcsec $\times$ 98\arcsec obtained by $7 \times 7$ binning of the beam, superimposed on the $^{13}$CO integrated intensity map. The map is overlaid with C$^{18}$O moment 0 contours (in red) at levels of 0.55, 0.92, 1.58, 2.77, and 3.69 K km s$^{-1}$. Each spectrum has been smoothened using a Gaussian filter with a $\sigma$ value of 1 km s$^{-1}$ for $^{13}$CO emission and 2 km s$^{-1}$ for C$^{18} $O emission to enhance the signal quality. For clearer visualization, the C$^{18}$O spectra have been scaled by amplifying the antenna temperature (T$_A$) by a factor of 2.}
\label{Fig: Spectral profiles of 16 regions}
\end{center}
\end{figure*}

Figures~\ref{Fig: Spectral profiles of 16 regions} presents the spectral profiles of 16 regions, each approximately 98$\arcsec$ $\times$ 98$\arcsec$ in size, extracted from the $^{13}$CO and C$^{18}$O emission maps. These regions, delineated by black boxes superimposed on the $^{13}$CO (3-2) integrated intensity map overlaid by C$^{18}$O (3-2) moment 0 contours, correspond to $7 \times 7$ binning of the JCMT-HARP observation’s 14" beam size. 

In the pursuit of identifying infall signatures, a critical criterion involves resolving double velocity peaks in the $^{13}$CO (3-2) spectra (in black), with the blue-shifted component displaying a higher intensity than the red-shifted component. Simultaneously, the C$^{18}$O (3-2) spectral profile (in blue) should exhibit a single velocity peak whose position coincides with the dip between the two peaks in the $^{13}$CO spectrum.

In Figure~\ref{Fig: Spectral profiles of 16 regions} for $^{13}$CO emission, in Region 3, two distinct peaks appear at velocities of 7.97 and 8.44 km s\(^{-1}\). Similarly, Region 6 reveals peaks at 6.90 and 7.68 km s\(^{-1}\), while Region 9 exhibits peaks at 6.74 and 8.10 km s\(^{-1}\). However, in these regions (3, 6, and 9), the red-shifted components are more prominent than their blue-shifted counterparts, suggesting the absence of gas infall motion.

Conversely, in $^{13}$CO emission in Region 4, the blue-shifted peak at 7.70 km s\(^{-1}\) is more intense than the red-shifted peak at 8.33 km s\(^{-1}\). A similar trend is observed in Region 7, where the blue-shifted peak at 7.75 km s\(^{-1}\) dominates over the red-shifted peak at 8.25 km s\(^{-1}\). However, the C$^{18}$O emission from Region 4 is deemed negligible, and therefore, not considered in the analysis. Furthermore, in the C$^{18}$O spectrum of Region 7, no single velocity peak aligns with the dip between the double peaks in the $^{13}$CO spectrum. This lack of alignment indicates the absence of infall motion in Region 7 which is proximal to the C2 clump. 

\section{Discussion}
\label{section: Discussion}

\subsection{Star Formation Potential and Comparative Analysis of Dense Clumps}
\label{section: Star Formation Potential of the Clumps}

Based on both the virial parameter and energy budget analyses, clump C1 appears to be gravitationally bound, clump C2 is unbound, and clump C3 is found to be in a state close to virial equilibrium. As discussed in Section~\ref{section: Identification of clumps}, clumps C1 and C2 exhibit clear signatures of star formation activity, while no such activity has been detected within C3. Although C2 hosts Class 0 protostars, the clump appears to be gravitationally unbound, likely due to feedback from protostellar outflows, as suggested by \citet{stanke2002triggered}, which may have injected sufficient energy to overcome the gravitational binding and suppress further collapse.

The C1 clump, on the other hand, contains mid-infrared sources \citep{stanke2002triggered} and pre-main-sequence stars \citep{alcala2004multi}, and remains gravitationally bound, reinforcing its role as an active and potentially ongoing site of star formation. Although clump C3 currently shows no observable signs of ongoing star formation, its near-virial equilibrium state suggests that it may evolve into an active star-forming region in the future, provided that the appropriate physical conditions—such as sufficient mass accumulation or external triggering—are met.

For a comparative perspective, we examine the physical properties of clumps identified by \citet{walker2013structure} in the NGC~2068 region, based on JCMT-HARP observations of C$^{18}$O (3--2) emission. NGC~2068 is a prominent reflection nebula embedded within the Orion~B molecular cloud, situated at a distance of approximately 415~pc \citep{walker2013structure}, and hosts a population of young stellar objects (YSOs) with estimated ages as young as $\sim$10$^{5}$ years \citep{strom1975m78}. In their analysis, the effective radii of the identified C$^{18}$O clumps span from 0.03 to 0.10~pc, while clump masses range between 0.7 and 11.9~M$_\odot$. The corresponding virial masses lie between 0.2 and 2.7~M$_\odot$, yielding virial parameters in the range 0.1 to 0.8—suggesting that all clumps are gravitationally bound. The one-dimensional velocity dispersions ($\sigma_{1\mathrm{D}}$), derived from the C$^{18}$O spectra at the peak emission positions, range from 0.16 to 0.42~km~s$^{-1}$, with a mean of approximately 0.28~km~s$^{-1}$, as detailed in their Table A2.

In contrast, our investigation of the L1616 region using the Clumpfind algorithm reveals the effective clump radii in the range of 0.08--0.13~pc, broadly consistent with the effective radii reported for NGC~2068. The masses of the L1616 clumps fall between 4.5 and 8.2~M$_\odot$, overlapping with the mass distribution found in NGC~2068. However, the one-dimensional velocity dispersions obtained from Gaussian fitting of the C$^{18}$O spectra at the peak C$^{18}$O emission (right panel of Figure ~\ref{Fig: Clump finding using Clumpfind algorithm})—measured as 0.26, 0.65, and 0.21~km~s$^{-1}$ for clumps C1, C2, and C3, respectively—yield a mean value of 0.37~km~s$^{-1}$, which is approximately 1.3 times higher than the mean reported for NGC~2068. Notably, the virial masses derived in our study, ranging from 9 to 21~M$_\odot$, are significantly greater than those reported by \citet{walker2013structure}, and the associated virial parameters lie in the range 1.3 to 2.7.

This marked discrepancy in virial mass and virial parameter indicates fundamental differences in the dynamical states of the clumps within the two regions. In NGC~2068, where YSOs are already present, the gravitational binding of clumps appears well established. Conversely, in L1616, only clump C1 satisfies the gravitational boundedness criterion. Clump C2, despite its association with the Class~0 protostars, appears gravitationally unbound, while clump C3—though devoid of current star formation activity—exhibits a virial state close to equilibrium. These variations likely reflect differences in the underlying physical conditions, evolutionary stages, and external environmental influences shaping the molecular cloud structures in NGC~2068 and L1616.

It is important to highlight that in the study by \citet{walker2013structure}, the clump masses derived from C$^{18}$O emission were estimated under the assumption of a uniform excitation temperature of 16.1~K for all clumps. This value was adopted from the mean excitation temperature of the entire NGC~2068 region, as derived by \citet{ikeda2009survey} using NH$_{3}$ transition measurements. Such a simplification inherently introduces some uncertainties in the mass estimates, as acknowledged by the authors. In contrast, our analysis accounts for spatial variation by estimating the clump masses using their respective excitation temperatures, which are found to be on average 16.7~K for clump C1, 14.3~K for C2, and 12.3~K for C3. These values demonstrate that the excitation temperatures within individual clumps in L1616 are broadly consistent with the mean excitation temperature reported for the entire NGC~2068 complex. Although the excitation temperatures reported by \citet{ikeda2009survey} are derived from NH$_{3}$ transitions, which typically trace denser gas than our $^{13}$CO-based measurements, the close agreement of these temperature values suggests a broadly similar thermal environment across the two regions despite the difference in molecular tracers employed.

\subsection{Shaping and Star Formation in the L1616 Cloud: Effects of Ionizing Radiation}
\label{section: Shaping and Star Formation in the L1616 Cloud: Effects of Ionizing Radiation}

As illustrated in the left panel of Figure~\ref{Fig: Planck 857 GHz image and WISE 12 micron image}, the ten ionizing stars originally identified by \citet{ramesh1995study} were subsequently adopted by \citet{saha2022magnetic} as the primary influencing sources potentially responsible for triggering star formation in L1616. In their analysis, \citet{saha2022magnetic} examined the dynamics of YSOs in the L1616 region \citep{alcala2004multi, stanke2002triggered, zari20183d, gandolfi2008star} using astrometric parameters from \textit{Gaia} Early Data Release~3 (EDR3; \citealt{brown2021gaia}), particularly the proper motions in right ascension ($\mu_{\alpha*}$) and declination ($\mu_{\delta}$), measured relative to these ten massive OB stars (see Figure B1 of \cite{saha2022magnetic}). Based on the classification by \citet{hernandez2005herbig}, stars~2 and~4 are associated with the Orion~OB1bc subgroup, whereas stars~7, 8, 9, and~10 belong to Orion~OB1a. Star~1 is a confirmed member of the $\sigma$~Orionis cluster, as cataloged in the Mayrit database \citep{caballero2008stars}, while star~3 is associated with $\epsilon$~Orionis \citep{lesh1968kinematics, saha2022magnetic}. Both $\sigma$~Orionis and $\epsilon$~Orionis are members of the Orion~OB1b subgroup, as suggested by \citet{voss2010probing}. Stars~5 and~6 are identified as members of Orion~OB1c, following the classification of \citet{simon2010chemical}. Consequently, the identified stars span the Orion~OB1a, OB1b, and OB1c subgroups, with none belonging to the Orion~OB1d association—known to represent a younger star-forming population compared to the other subgroups \citep{bally2008overview}. Among these, $\epsilon$~Orionis (B0\,Ia; star~3) emerges as the principal influencing source for both the morphology and star formation activity in the L1616 cloud, as the peak CO emission is spatially oriented toward the direction of its radiation field, an interpretation originally proposed by \citet{ramesh1995study} and later supported by \citet{saha2022magnetic}.

The incident far-ultraviolet (FUV) radiation field impinging upon the L1616 globule, expressed in units of the Habing field G$_{0}$ ($1.6 \times 10^{-3}$ erg cm$^{-2}$ s$^{-1}$) \citep{habing1968interstellar}, was estimated using the formulation provided by \citet{codella2001star} and originally derived by \citet{sternberg1989infrared}:

\begin{equation}
    G_{0} = 170 \, \frac{L}{L_{\odot}} \left( \frac{d}{1 \, \mathrm{pc}} \right)^{-2} T_4^{-4} \left[ \exp\left( \frac{14.4}{T_4} \right) - 1 \right]^{-1},
    \label{eq: G0 equation}
\end{equation}
where $L$ is the stellar luminosity, $d$ is the projected distance between the ionizing star \citep{saha2022magnetic} and the globule, and $T_{4}$ is the effective stellar temperature normalized to $10^{4}$ K. Luminosities and temperatures for the stellar sources contributing to the FUV field were adopted from Table 1 of \citet{panagia1973some}, as indicated in Figure~\ref{Fig: Planck 857 GHz image and WISE 12 micron image}. However, for certain spectral types (B2.5V, B1.5V, and O9.7V), the relevant stellar parameters are not listed. In such cases, we adopted interpolated values by averaging the luminosities and temperatures of adjacent spectral subtypes: B2 and B3 for B2.5V, B1 and B2 for B1.5V, and O9 and B0 for O9.7V. Additionally, a star denoted as B2 in \citet{saha2022magnetic}, lacking a specific luminosity class, was identified in the SIMBAD database as type B2 IV/V.

Applying equation~\ref{eq: G0 equation} to our sample, we derive a cumulative FUV flux incident on L1616 of $G_{0} \approx 3.73$. Notably, $\epsilon$~Orionis alone contributes an FUV flux of approximately 2.65, while the remaining stars individually provide only $\sim$0.6--19\% of this value. The highest secondary contribution (19\%) arises from the O9.7\,V star (star~6). This FUV flux distribution strongly supports the conclusions of \citet{ramesh1995study} and \citet{saha2022magnetic}, reinforcing the interpretation that $\epsilon$~Orionis is the dominant ionizing source driving the PDR L1616. This derived total FUV flux is consistent with modestly irradiated environments: the interstellar radiation field spans from $G_{0} \approx 1.7$ in quiescent regions up to $G_{0} \gtrsim 10^{6}$ in extreme environments near luminous O-type stars within 0.1 pc of dense molecular material \citep{hollenbach1997dense}. The estimated flux implies that the L1616 globule lies within a moderate PDR influenced by nearby OB-type stars.

To further characterize the depth of FUV penetration into the molecular cloud, we estimate the visual extinction, $A_V$, across the C$^{18}$O (3--2) emission region using the canonical relation \citep{bohlin1978survey, rawat2023probing}:

\begin{equation}
    A_V = \frac{N(\mathrm{H}_2)}{9.4 \times 10^{20} \, \mathrm{cm}^{-2} \, \mathrm{mag}^{-1}},
    \label{eq: AV equation}
\end{equation}
yielding an average extinction of $A_V \approx 9.7$. This extinction is in excellent agreement with the threshold reported by \citet{hollenbach1997dense}, who modeled static photon-dominated regions and demonstrated that the transition from atomic to molecular oxygen gas typically occurs at $A_V \approx 10$. Beyond this layer, the attenuation of FUV photons becomes significant. Hence, our derived $A_V$ and $G_0$ values together confirm that L1616 resides in a well-developed PDR, shaped by the radiation from nearby OB stars in the Orion complex.

Our JCMT-HARP observations, which probe deeper into this PDR, reveal three prominent clumps—C1, C2, and C3—embedded within the C$^{18}$O (3--2) emission region (see Figure~\ref{Fig: Clump finding using Clumpfind algorithm}). Notably, the mid-infrared (MIR) sources associated with clump C1 exhibit stellar ages of the order of $10^6$ yr, consistent with the evolutionary age of the nearby star HD 293815 \citep{stanke2002triggered}. In contrast, clump C2 hosts Class 0 protostars, representing an earlier stage of stellar evolution, thereby suggesting a temporal gradient in star formation activity across the cloud.

The L1616 globule is predominantly shaped by the radiative and mechanical influence of massive stars belonging to the Orion OB1a, OB1b, and OB1c subgroups \citep{saha2022magnetic}. According to \citet{bally2008overview}, the OB1a subgroup comprises stars with ages of 8--12 Myr, OB1b members have age between 1.7--8 Myr, and OB1c members range between 2--6 Myr. These chronological constraints imply that the onset of star formation within L1616 is a relatively recent event, likely triggered by the radiative influence of older, nearby OB stars. The age difference between the C1 and C2 clumps also supports a sequential star formation scenario, with C1 preceding C2 \citep{stanke2002triggered}.

Regarding the mechanism responsible for triggering this star formation episode, several lines of evidence disfavor the possibility of a supernova-driven origin. \citet{stanke2002triggered} argue that supernova shocks propagating through the Sco-Cen association would likely traverse low-density intercloud media $ (< 10^{3} \ \mathrm{cm}^{\bm{-3}})$, rendering them inefficient at compressing molecular clouds. Furthermore, the shock velocities in such media are typically a few tenths of km s\(^{-1}\), far below the 10 km s\(^{-1}\) required to induce collapse in dense cores, as discussed by \citet{preibisch1999history}.

Instead, we propose that radiation-driven implosion (RDI) is the dominant mechanism behind the initiation of star formation in L1616, consistent with the interpretations put forth by \citet{stanke2002triggered} and \citet{saha2022magnetic}. In this scenario, pre-existing dense clumps are compressed by ionizing FUV radiation, resulting in triggered gravitational collapse. This interpretation is supported by both the morphology and age structure of the region. The observed clumps (C1, C2) lie along an east-to-west progression that is consistent with sequential triggering, with C1—the easternmost and closest to the ionizing OB stars—forming first. The absence of evident star formation in C3 suggests it has not yet received sufficient external pressure to overcome internal support mechanisms.

The cometary morphology of L1616 can be understood as a consequence of the interplay between RDI and the rocket effect. As described by \citet{saha2022magnetic}, RDI leads to the compression and eventual equilibrium of dense clumps with the surrounding ionized medium, while the rocket effect results in the acceleration of the cloud material away from the ionizing sources, thereby producing an elongated, lower-density tail-like structure. This combination of physical processes elegantly accounts for the cloud's morphology, the sequential nature of star formation, and the presence of embedded clumps at varying evolutionary stages.

Despite the modest FUV radiation field, L1616 demonstrates that radiative feedback can still drive clump evolution and trigger collapse when pre-existing density enhancements are present. Even a weak but sustained FUV field can induce pressure imbalances, especially in regions that are already gravitationally marginal. The triggering here is not primarily due to high radiation intensity but rather to the susceptibility of dense clumps to compression.

\section{Summary}
\label{section: Summary}

This study explores the gas kinematics and physical characteristics of the cometary globule L1616, a prominent photodissociation region, with the aim of uncovering the mechanisms driving star formation within it. In this study, we report for the first time the identification of three distinct clumps that are actively involved in star formation processes. By estimating the velocity dispersion, non-thermal velocity dispersion, and Mach number for the identified clumps—C1, C2, and C3—within the L1616 PDR, we find compelling evidence of dynamic activity, particularly in clump C2. The elevated Mach number observed in C2, which harbors Class 0 protostars as reported by \citet{stanke2002triggered}, suggests the presence of outflow activity. However, due to the absence of $^{12}$CO (3--2) data, we are unable to conclusively identify the characteristic outflow wing features in the spectral profiles shown in Figure~\ref{Fig: Spectral profiles of 16 regions}.

We assess the gravitational stability of the clumps through both virial parameter and energy budget analyses, which collectively indicate that clump C1 is gravitationally bound and clump C3 is in a state of near-virial equilibrium. Clumps C1 and C2 currently exhibit signs of active star formation, while C3—though presently quiescent—appears well-positioned to develop into a star-forming core in the near future. In contrast, despite the presence of YSOs within clump C2, its gravitationally unbound nature suggests that it is unlikely to sustain further star formation activity going forward.

The estimated FUV flux incident on L1616, derived from stellar sources identified in Figure~\ref{Fig: Planck 857 GHz image and WISE 12 micron image}, is not sufficiently intense to independently trigger the formation of dense clumps. Therefore, we propose that the clumps identified in this study were pre-existing structures within the L1616 PDR. Such clumpy substructures are commonly observed in photon-dominated regions, as noted by \citet{hollenbach1997dense}. In this scenario, radiation-driven implosion acts upon these pre-existing overdensities, enhancing their compression and driving them toward gravitational instability, ultimately initiating star formation. Furthermore, the sequential nature of star formation observed in L1616 is consistent with the progressive action of RDI across the region. The combined influence of RDI and the rocket effect, as discussed by \citet{saha2022magnetic}, likely contributes to the characteristic cometary morphology of the L1616 globule.

\section*{Acknowledgements}
We are grateful to the anonymous referee for the thoughtful and detailed feedback provided during the review process. Their insights have helped us to enhance the manuscript.

We thank the staff of the James Clerk Maxwell Telescope (JCMT) for their support during the observations and acknowledge the JCMT HARP data used in this study, obtained under project code E23BU009. The JCMT is operated by the East Asian Observatory on behalf of its partners: the National Astronomical Observatory of China, the National Astronomical Observatory of Japan, the Korea Astronomy and Space Science Institute (KASI), the Academia Sinica Institute of Astronomy and Astrophysics of Taiwan, and the Science and Technology Facilities Council of the United Kingdom.

Additionally, we utilized the WISE 12 and 4 micron emission data, which are publicly available from NASA's SkyView Virtual Observatory (https://skyview.gsfc.nasa.gov).

This research was supported by the Indian Institute of Astrophysics (IIA), funded by the Department of Science and Technology (DST), Government of India. E.J.C. was supported by the Basic Science Research Program through the National Research Foundation of Korea (NRF) funded by the Ministry of Education (grant number NRF-2022R1I1A1A01053862). C.W.L. was supported by the Basic Science Research Program through the National Research Foundation of Korea (NRF) funded by the Ministry of Education, Science and Technology (NRF-2019R1A2C1010851), and by the Korea Astronomy and Space Science Institute grant funded by the Korean government (MSIT; project No. 2022-1-840-05).

\section*{Data Availability}
The observational data used in this study were obtained from the James Clerk Maxwell Telescope (JCMT), specifically using the HARP instrument for the \(^{13}\text{CO}\) and C$^{18}$O (3-2) emission lines. The JCMT data can be accessed through the JCMT Science Archive. Additional data from the Wide-field Infrared Survey Explorer (WISE) were obtained via the NASA SkyView virtual observatory.

The data analysis was carried out using Python, with the following libraries and tools: Astropy for handling astronomical data \citep{price2018astropy}, SciPy for numerical analysis \citep{virtanen2020scipy}, NumPy for array operations \citep{harris2020array}, and Clumpfind for identifying clumpy structures in the emission data \citep{williams1994determining}.



\bibliographystyle{mnras}
\bibliography{refL328} 







\bsp	
\label{lastpage}
\end{document}